\documentclass[12pt,preprint]{aastex}

\newcommand{\myemail}{naomi@fcrao.umass.edu}

\newcommand{\Msolar}{\mbox{\,M$_\odot$}}        % solar mass
\newcommand{\Lsolar}{\mbox{\,L$_\odot$}}        % solar luminosity
\newcommand{\kms}{\mbox{\,km\,s$^{-1}$}}                           % Kms-1
\newcommand{\Ta}{T$_{_{\rm A}}\!\!\!$*}                 % Ta* (corr. ant. temp)
\newcommand{\co}{$^{12}$CO}                             % 12co
\newcommand{\xco}{$^{13}$CO}                            % 13co
\newcommand{\xxco}{C$^{18}$O}                           % c18o
\newcommand{\degrees}{$^{\circ}$}
\newcommand{\x}{\times}
\defcitealias{fuente02}{FMBRP}

\shorttitle{Molecular Gas around Young Stellar Clusters}
\shortauthors{Ridge et al.\ }

\begin{document}

\title{A \xco\ and \xxco\ Survey of the Molecular Gas Around Young
Stellar Clusters Within 1\,kpc of the Sun}

\author{Naomi A.\ Ridge}
\affil{Five College Radio Astronomy Observatory, Dept. of Astronomy, University of Massachusetts, 
Amherst, MA 01003, USA}
\email{\myemail}

\author{T.L. Wilson}
\affil{Submillimeter Telescope Observatory, University of Arizona, Tucson, AZ 85721, USA \& 
Max Planck Institut f\"ur Radioastronomie,
Auf dem H\"ugel 69,
53121 Bonn,
Germany
}

\author{S.T. Megeath, L.E. Allen \& P.C. Myers}
\affil{Harvard-Smithsonian Center for Astrophysics, 60 Garden Street, Cambridge, MA 02138, USA}
%\email{}

%%ABSTRACT
\begin{abstract}

As the first step of a multi-wavelength investigation into the
relationship between young stellar clusters and their environment we
present fully-sampled maps in the J=1--0 lines of \xco\ and \xxco\ and
the J=2--1 line of \xxco\ for a selected group of thirty young stellar
groups and clusters within 1\,kpc of the Sun.  This is the first
systematic survey of these regions to date.  The clusters range in
size from several stars to a few hundred stars.  Thirty fields ranging
in size from $8'\times 8'$ to $30'\times 60'$ were mapped with 47$''$
resolution simultaneously in the two J=1--0 lines at the Five College
Radio Astronomy Observatory.  Seventeen sources were mapped over
fields ranging in size from $3'\times 3'$ to $13'\times 13'$ in the
J=2--1 line with 35$''$ resolution at the Submillimeter Telescope
Observatory. We compare the cloud properties derived from each of the
three tracers in order to better understand systematic uncertainties
in determining masses and linewidths.  Cloud masses are determined
independently using the \xco\ and \xxco\ transitions; these masses
range from 30 to 4000~\Msolar.  Finally, we present a simple
morphological classification scheme which may serve as a rough
indicator of cloud evolution.
\end{abstract}


\section{Introduction}
\label{intro}

It is well established that most stars form in clusters. For example,
in an analysis of the 2MASS second incremental release catalogue of
the Orion A, Orion B, Perseus and Monoceros molecular clouds,
\citet{carp2000} found a total of 1200 isolated stars and $\sim$3000
stars in fourteen clusters, six of these clusters with more than 100
members each.  \citet{hodapp94} performed a K$'$-band survey of 165
regions with known molecular outflows, and found that one third of the
outflows were located within a cluster of stars.  Clustered star
formation spans an enormous range in the number of stars from small
groups of several stars in the Taurus dark clouds to the largest known
young cluster in our galaxy, NGC\,3603, containing many thousands of
stars and finally to proto-globular clusters containing up to one
million stars. A complete understanding of star formation therefore
requires a theory which describes the full range of star formation
from small groups to globular clusters.

The formation of stars in clusters is thought to occur over a
$\sim$1\,Myr time span \citep[e.g.][]{h97,ps2000}, during which the
material to form stars is continually drawn from a reservoir of
molecular gas. It is from this reservoir, through a process not fully
understood, that fragments condense out of the gas and collapse to
form stars.  It is therefore likely that the rate of star formation
and the number of stars which ultimately form depend on the properties
of the molecular reservoir.  Previous studies have observed a clear
difference between the molecular gas in regions containing small
groups of stars, such as Taurus, and regions containing large
clusters, such as Orion
\citep{jma99,onishi96,lbs91,lada92,css90,css95,tmwr98}.  A systematic
study of the molecular gas in clusters spanning a range in the number
and density of constituent stars can probe in detail how the
properties of the parental cluster-forming gas dictates the properties
of the forming clusters.

Such a systematic survey could also provide a better picture of the
evolution of the cluster-forming molecular gas.  The young stars are
thought to dissipate the parental cluster-forming gas through winds
and radiation, resulting in the termination of ongoing star formation;
however, the timescale and exact mechanism of the dissipation remains
poorly understood.  The process of gas dissipation has been studied in
the vicinity of Herbig Ae/Be stars by \citet[][ hereafter
\citetalias{fuente02}]{fuente98, fuente02}, most of which are members
of clusters.  While these observations provide insight into the
mechanism and timescale for the dissipation of gas around individual
stars, the dissipation of the extended cluster-forming gas remains to
be studied in detail.

We have therefore begun a program of multi-wavelength observations
which, when combined with existing infrared and optical data, will
yield detailed information about both the stars and their surrounding
gas in a large sample of young stellar groups and clusters within
1\,kpc of the sun. The final data set will include sensitive ground
based wide-field optical and near-infrared imaging and spectroscopy,
millimeter spectral-line and continuum maps and mid-infrared
observations obtained with the Space Infrared Telescope Facility
(SIRTF). Once complete, it will be made available to the community via
a web-based database, providing an invaluable resource for future
star-formation studies.

In this paper we present our \xco\ and \xxco\ molecular line
data. \xxco\ is an excellent tracer of the column density in the warm,
dense gas typical of cluster forming regions \citep*{gbl97}. Although
the freeze out of \xxco\ has been observed in cold, dark clouds
\citep{bahl02}, this should only affect small pockets of cold, dense
gas within each cluster, such as gravitationally unstable fragments of
gas undergoing collapse.  The overall structure of the warm gas
pervading the cluster is expected to be well traced by \xxco.  In
addition to tracing the structure of the molecular gas, the \xxco\
lineshapes and velocities are good tracers of the gas kinematics and
are relatively unaffected by optical depth effects.  Although emission
from the more abundant \xco\ can be optically thick toward the cluster
center, the \xco\ emission can be traced further into the outer, more
tenuous regions of the cluster where \xxco\ is too weak to be mapped
efficiently. These data therefore enable us to characterise the
cluster-forming molecular gas from the regions of peak density toward
the centers of each cluster out to the non-star-forming gas of the
surrounding molecular cloud. Here we concentrate on examining general
trends in the morphological properties of the cluster gas. Detailed
analysis of the gas kinematics in each of the individual sources will
be presented in subsequent papers (Ridge et al.\ in prep.).

Section \ref{sample} describes how our sources were selected, section
\ref{obs} gives details of the observing and data reduction
procedures. An ``atlas'' of the clusters is presented and cluster
properties are compared in section \ref{res}. Finally we summarise our
main conclusions in section \ref{summ}.

\section{The Sample}
\label{sample}
To provide a rich sample of sources which are relatively nearby and
thus can be observed with high angular resolution and sensitivity, we
have compiled from the literature a list of all known young stellar
clusters within 1\,kpc of the Sun (Christopher et al.\ in prep.).
This list contains a remarkable diversity of regions from small groups
with several stars to dense clusters containing hundreds of stars.  We
use the term ``cluster'' in this paper to refer to both small groups with
five or more stars, and large clusters of several hundred stars.  Only
these large clusters may eventually form bound open clusters after the
dissipation of their parental molecular gas, while most of the smaller
systems should quickly disperse \citep{am2001}.  Here we present the
subset of regions that are easily observable from the north.  Clusters
in the regions of Orion and Ophiuchus are not included as they are
well studied in the literature. A full list of the objects in the
survey, their distances, and the co-ordinates we adopted as the map
center for each is given in table \ref{sourcetab}. Where available in
the literature the number of stars in the cluster is also given, and a
brief summary of previous observations of each source is included in
appendix \ref{sources}.

The IRAS point source catalogue was searched for objects within 3$'$
of each of the map centers. Where no source could be located within
this radius, the search radius was increased in 30$''$ increments
until at least one IRAS point source was found. Far-infrared (FIR)
luminosities were calculated from the 12--100\micron\ IRAS fluxes
using the following formula \citep{cas86}:
\begin{equation}
\frac{L_{\rm FIR}}{L_{\odot}} = 4.7\times10^{-6}D^2 
\left(
\frac{S_{12}}{0.79} +
\frac{S_{25}}{2} +
\frac{S_{60}}{3.9} +
\frac{S_{100}}{9.9}
\right), %L_{\odot}
\end{equation}
where D is the distance in pc and $S_{\lambda}$ is the flux density in
Jy. These luminosities are also listed in table \ref{sourcetab}. If
more than one IRAS source was found within the search area then the
luminosity of the brightest source is listed.  We will use the FIR
luminosity in this paper as an indicator of the size (mass) of the
stellar component of the clusters, but for more evolved clusters, such
as IC\,348, where most of the stars are optically visible this
assumption is not valid.

Figure \ref{dist}a shows a histogram of the number of sources with
distance. The solid line represents the full sample and the dashed
line indicates those sources which were observed at both the Five
College Radio Astronomy Observatory (FCRAO) and the Submillimeter
Telescope Observatory (SMTO).  The histograms show the wide range of
distances toward our sample regions, which spans a factor of seven
from the nearby L\,1551 dark cloud to the distant GGD\,4 cluster.
This translates into a factor of seven in spatial resolution.  The
biases this may introduce in a comparative analysis of the clusters
are mitigated by the range of cluster properties sampled at a given
distance, as shown in Figure \ref{dist}b.  Although there is a trend
of increasing FIR luminosity with distance, the upper and lower
envelope of this trend differ by a factor of one hundred in
luminosity.  Futhermore, biases due to distance can be eliminated by
comparing regions in the same molecular cloud.  In particular, we have
two or more clusters within the Perseus cloud, NGC\,1499 cloud,
Monoceros cloud, the Cepheus cloud, and the Cepheus Bubble.  By
studying clusters within a single cloud or cloud complex, we can
undertake comparative studies which are unaffected by variations in
spatial resolution, spatial coverage, or sensitivity.

\section{Observations and Data Reduction}
\label{obs}

\subsection{FCRAO Observations}

Observations in the \xco~1--0 (110.201\,GHz) and \xxco~1--0
(109.782\,GHz) transitions were carried out during several periods
between 2001 December and 2002 November at the FCRAO 14m telescope in
New Salem, Massachusetts, mostly during the commissioning stage of the
On-the-Fly (OTF) mapping technique.  OTF maps of each of the clusters
were made using the SEQUOIA 16-element focal plane array and a dual-IF
narrow-band digital correlator, enabling maps in \xco\ and \xxco\ to
be obtained simultaneously. Observations made after 2002 March
utilised the upgraded SEQUOIA-2 array with 32 elements.  The
correlator was used in a mode which provided a total bandwidth of
25\,MHz, with 1024 channels yielding an effective velocity resolution
of 0.07\kms. For a few objects (see table \ref{obstab}) data was
obtained in the 50\,MHz bandwidth mode, with a velocity resolution of
0.13\kms.  The weather was exceptionally stable during the runs with
system temperatures at 110\,GHz generally between 200 and 400\,K
(single sideband).

Maps were obtained by scanning in the RA direction, and an ``off''
source reference scan, was obtained after every two
rows. Off-positions were checked to be free of emission by performing
a single position-switched observation with an additional 30$'$
offset.  Map sizes are given in table \ref{obstab}. A data-transfer
rate (DTR) of either 1 or 2\,Hz was used, and in some cases the full
size maps were built up from several smaller submaps. When a DTR of
2\,Hz was used, maps were repeated to increase
signal-to-noise. Calibration was by the chopper wheel technique
\citep{ku81}, yielding spectra with units of \Ta. 
Pointing was checked regularly and found to vary by less
than 5$''$ rms.

The OTF technique was implemented at FCRAO in order to compensate for
the removal of the dewar rotation system in the summer of 2001. The
maps obtained are therefore not evenly sampled and a convolution and
regridding algorithm has to be applied to the data in order to obtain
spectra on a regularly sampled grid. This process was carried out
using software provided by the observatory \citep*{otfmanual}. Using
this software the individual spectra have a linear baseline
subtracted, then are convolved onto a regular 25$''$ grid weighted by
1/rms$^2$, yielding a Nyquist sampled map, and corrected for the main
beam efficiency of the telescope, which is 0.48 at 110\,GHz.

After regridding, the individual spectra had an rms sensitivity of
$\sim$0.2\,K per 25\,kHz channel (see table \ref{obstab}), with the
OTF technique yielding maps with extremely uniform
sensitivity. Additionally, the simultaneous observation of the two
lines gives perfect registration between the
\xco\ and \xxco\ maps.

\subsection{SMTO Observations}

Observations in the \xxco~2--1 (219.560\,GHz) transition were carried
out at the Heinrich-Hertz Telescope (HHT) during several periods in
the winter and spring of 2001--2002.  The receiver was a single
channel SIS mixer with a double sideband receiver noise temperature of
about 120\,K, depending on receiver tuning. The total (single
sideband) system noise temperature was between 270 and 500\,K
depending on weather. The telescope beamsize was 35$''$ at the line
frequency. The pointing was checked on well known calibrators or
planets, and found to be 2$''$ rms.  The main beam efficiency was
measured to be 0.78 at the line frequency.  However, because of 30\%
variations in the sideband ratios, the results were scaled to
measurements of standard regions such as the Orion\,KL nebula. Spectra
have units of \Ta.

The data were taken employing OTF mapping. The procedure was first to
perform a hot-sky calibration using a chopper wheel, then to map a
2$'$ by 2$'$ or 2.5$'$ by 2.5$'$ region. These regions were mapped by
scanning in R.A. with a 15$''$ separation between the Declination
rows. A reference region either 10$'$ or 15$'$ lower in R.A. and
offset from the center of the map, was used. A measurement on the
reference position, which had been previously checked for the presence
of emission, was taken before measuring each row.  The raw OTF maps
were convolved with a Gaussian beam and regridded onto a regular
15$''$ grid in the {\em GILDAS} data reduction environment. The final
images consist of mosaics of the individual OTF maps.  After the
regridding process, individual spectra had an rms sensitivity of
$\la$0.17\,K per 250\,kHz (=0.34\,\kms) channel.

\section{Results}
\label{res}

Integrated intensity maps of the \xco~1--0 and \xxco~1--0 emission are
presented in figures \ref{cont0} through
\ref{cont11}. The \xco~1--0 maps have contours at 10\% to 100\%  of
the maximum integrated intensity with intervals of 10\%, except where
indicated otherwise in the caption. The base contour level is $\gg
3\sigma$. The \xxco~1--0 maps have a base contour level of 3$\sigma$,
with contours at 1$\sigma$ intervals in most cases. In a few sources
different contour levels are used, and are described in the captions.
Integrated intensity maps of the \xxco~2--1 emission are presented in
figures \ref{smto0} through \ref{smto3}, with contours at 10\% to
100\% of the maximum integrated intensity.  The value of the maximum
integrated intensity in each map is given in the top right corner of
the figures for reference. We have indicated the positions of known
stellar or protostellar sources in the regions by triangles (embedded
sources) or stars (optically-visible sources).  The \xxco\ 1--0
emission was too weak to be detected in only two sources, HD\,216629
and VV\,Ser.  This was particularly surprising in HD\,216629 where the
\xco\ emission was relatively strong at the peak position.

\subsection{Morphology}
The clusters show a range of morphologies, from compact, roughly
spherical (e.g.\ CB34), cometary (e.g.\ S140) to diffuse, extended
emission (e.g.~VVSer). Based on our examination of the \xco\ and
\xxco\ maps, we have assigned each source a morphological
description, which is listed in table \ref{classtab}. Some of the
sources could be placed in more than one category, for instance
NGC\,7129 shows both a compact core and a cavity.  Many of the sources
show cores surrounded by more extended envelopes. The emission rarely
exhibits circular symmetry; the cores are usually somewhat elongated,
and their envelopes are often more elongated than the cores, possibly
because the larger envelopes are better resolved.  In some cases an
elongated core in \xco\ appears as a chain of multiple cores when
observed in \xxco\, which is less affected by optical depth (e.g. IC
348).  In many cases, multiple peaks are apparent; we have tabulated
the number of ``significant peaks'' (N$_{peaks}$) detected in \xco\
for each region in table \ref{classtab}.  We have used the
\xco\ data to identify peaks due to the higher signal to noise in this
tracer; however, in several cases, additional peaks are apparent in
the \xxco\ maps due to the lower optical depth in the \xxco\ lines.

In many of the clusters, the most luminous members have been
identified, either through the IRAS point source catalog, or through
catalogs of Herbig Ae/Be stars \citep{testi97,testi98}.  As an
indicator of whether the observed \xco\ peaks contain known sites of
recent or ongoing star formation, the distance (d$_*$) between the
embedded source or star and its nearest \xco\ peak is shown for each
source in table
\ref{classtab}.  In cases where there are multiple peaks and/or stars,
the smallest distance is reported.

The observed structures in the \xco\ and \xxco\ maps are most likely a
combination of features due to initial conditions in the cloud and
interaction with already formed stars. The compact cores, such as
GGD\,4 and CB\,34 are likely to be still close to their initial state,
while clouds with cavities such as HD\,200775 are breaking apart due
to star-cloud interactions.  Most of the Herbig Ae/Be type stars in
the sample appear to be located in regions which have morphologies
consistent with dispersal of their surrounding molecular gas, while
the embedded sources appear to be preferentially located in gas which
is more centrally condensed. Nineteen of the thirty regions have
\xco\ peaks coincident with or close to a stellar or IRAS position. All
but two of these display a centrally condensed morphology (core or
core + envelope). We therefore interpret the proximity of the stellar
source to the map peak as an indicator of youth of the region.  Based
on these interpretations we propose a sequence of morphologies, or
``developmental classes'' as follows:

\begin{description}
\item[I] A single significant peak in \xco\ and \xxco\ indicative of
one dominant molecular core.  An extended envelope is commonly detected
around the core.  An embedded source is found within the
50\% contour of the peak \xxco\ emission, e.g.\ CB\,34.  Eleven of the
clusters fall into this class, including a subset of five
bright-rimmed globules, noted by ``BRG'' in table \ref{classtab}.
The BRGs all show ``tails'' of extended \xco\ emission.

\item[II] The \xco\ and \xxco\ emission is distributed in multiple
peaks and/or elongated filaments, often with an extended envelope. An
embedded source falls within the 50\% contour of a \xxco\ peak or
filament, e.g. S\,171. In some cases with two known associated sources
(e.g. NGC 7129), one of the sources is located outside of peaks.  Ten
sources fall into this class.

\item[III] Known stars fall well outside of \xco\ and \xxco\ peaks and
filaments, e.g. HD\,200775.  Typically show diffuse, extended
emission, multiple \xco\ peaks, and multiple or no \xxco\ peaks.  In
one region, IC348, two of the three associated sources fall outside the
\xco\ emission. Nine sources fall into Class III.

\end{description}

The classifications are listed in table \ref{classtab}, and figures
\ref{cont0} through \ref{cont8} are grouped according to
classification. Since many of the Class I and III sources show similar
FIR luminosities, it is likely that the differing morphologies
reflect an evolutionary progression from Class I to Class III driven
by the disruption of the gas by the embedded stars.  Due to their
morphological complexity, Class II sources are more problematic.  Many
of these regions may be coeval with Class I sources, indicating
different initial conditions, while others showing distinct signs of
gas dispersion (e.g. NGC\,7129) may be in an intermediate evolutionary
state between Class I and III.

This scheme differs from the the classification scheme of
\citet{fuente98,fuente02}, although the \citeauthor{fuente02} sample
contains several sources in common with this
work. \citeauthor{fuente02} mapped small regions ($\sim$1\,pc) around
each of their sources in order to determine the molecular gas
distribution close to the Herbig Ae/Be stars, while in contrast, our
proposed classification is based on the global morphology of the
cluster forming cloud.  However, of the six sources common to both
\citeauthor{fuente02} and this work (MWC\,297, HD\,200775, HD\,216629,
VV\,Ser, BD+65\degrees1637 and LkH$\alpha$234\footnote{Both these last
two fall within the region of NGC\,7129 we observed}), both
classification schemes place these sources as the most evolved regions
of the sample, with one exception, NGC\,7129, but this is due to our
treatment of the region as a whole, while
\citeauthor{fuente02} considered the two stellar sources separately.
We see significant molecular gas emission away from the location of
the star, and outside the region covered by
\citeauthor{fuente02} in most cases.

\subsection{Masses}

The core masses determined from the three isotopomers/transitions are
given in table \ref{masstab}. Gas masses were determined from the
\xco\ and \xxco\ \hbox{1--0} transitions using a local thermal
equilibrium (LTE) approximation, assuming an excitation temperature of
20\,K and the distances given in table \ref{sourcetab}. This method
has been found to be accurate to within a factor of 2--4 \citep{rw00}
as long as the actual excitation temperature is less than 30\,K. Such
high excitation temperatures are expected only toward the embedded OB
stars that occupy a small fraction of the total cloud area in any of
these regions.  The \xxco~2--1 masses were determined by
calculating the column density of H$_2$ from the large velocity
gradient (LVG) relation:
\begin{equation}
N_{\rm H_2} = 2.65\times 10^{21} \int T_{\rm MB}(C^{18}O, 
J= 2\rightarrow1){\rm d} v,
\end{equation}
where $v$ has units of \kms, $T_{\rm MB}$ of Kelvins and $N_{\rm H_2}$
is given in cm$^{-2}$ \citep{rw00}. $N_{\rm H_2}$ can then be
converted to a total gas mass by assuming a distance (as given in
table \ref{sourcetab}), the mass of the hydrogen molecule and a helium
abundance.  For comparison, we have also calculated the LTE mass from
the \xxco~1--0 emission for the smaller area covered by the \xxco~2--1
observations (hereafter referred to as the central \xxco\ mass) and
this is given in column 5 of table \ref{masstab}.

We compare the masses derived from each of the three transitions in
figure \ref{mass1}, in order to investigate how the choice of tracer
and map size will affect our results. Figure \ref{mass1}b shows the
central \xxco~2--1 mass plotted against the central \xxco~1--0 mass.
The central mass determinations for the two \xxco\ transitions are
very consistent, indicated by the dashed line showing the 1:1
relation.  Just two sources, GGD\,12-15 and CepC fall significantly
from this relation.  This plot demonstrates that mass estimates from
different transitions using different approximations (LTE vs LVG) lead
to equivalent results.  Comparison of the central \xxco~2--1 masses,
measured with the SMTO, and the total \xxco~1--0 masses in the larger
region covered by the FCRAO observations (fig.\ref{mass1}c) indicates
that a significant fraction of the gas ($\sim$65\%) is found in
emission that extends outside the smaller \xxco~2--1 maps.
Nevertheless, a clear correlation is evident, suggesting that the mass
in the extended component grows linearly with the mass in the central
region.

The correlation between the envelope mass and the central mass is also
apparent in the strong correlation ($\sim 9\sigma$) between the \xco\
and \xxco~1--0 masses (Figure \ref{mass1}a).  However, the \xco\
masses are approximately a factor of two larger than the masses
determined from the \xxco~1--0 emission. This is indicated by the
solid line on figure \ref{mass1}a which is a least-squares-fit (LSF)
to the data, with a slope of 0.58$\pm$0.05.  The \xco\ traces lower
density gas since the abundance is 5 to 10 times higher than
\xxco\ and also because photon trapping allows the \xco\ to be excited
at a lower density than \xxco. In addition, the J=1--0 line is more
easily excited than the J=2--1 line because the Einstein A coefficient
is about a factor of 8 smaller.  Therefore the \xco\ emission is
detectable in more extended regions, where the \xxco\ emission is too
weak to detect, and is likely to be mostly due to gas in the envelope
which is too diffuse to detect in the weaker \xxco\ line.  The
relationship between the 'central' (\xxco~2--1) and 'total'
(\xco~1--0) masses shows a strong correlation ($\sim 9\sigma$) with
small scatter (Figure \ref{mass1}d).  This is surprising considering
the varied morphologies of the sources and the fact that the physical
size of the region mapped varies with distance.  The line represents a
LSF with a slope of 0.23$\pm$0.02, indicating that the central mass is
usually $\sim$ 1/4 of the total mass.

Histograms of the \xco~1--0 masses are presented in figure
\ref{mass_hist}. These show that 2/3 of the clouds are below
1000\,\Msolar\ when measured with \xco, and that 1/3 of the clouds are
below 500\,\Msolar. Figure \ref{mass_plots}a shows the \xxco~1--0
masses as a function of distance. Although there is an apparent trend
of higher mass with distance, the upper and lower envelope of this
trend differ by more than an order of magnitude.  This wide range of
masses present at any distance interval will minimise biases due to
distance in subsequent comparisons of cloud vs. cluster
properties. Finally, the \xxco~1--0 mass is plotted against the FIR
luminosity in figure \ref{mass_plots}b. The correlation here ($\sim
4\sigma$), indicating that more massive cores are forming more
luminous (massive) stars, is much stronger than any correlation
between either mass or FIR luminosity and distance (both $\la2\sigma$)
and therefore this effect is likely to be real.

\subsection{Sizes and Linewidths}
\label{s-l}

The sizes and linewidths of the cores are listed in Table \ref{lwtab}.
We define core size, R, here as $\sqrt{A/\pi}$ where $A$ is the total
area in pc$^2$ (i.e.\ number of pixels $\times$ pixel size in pc)
where the integrated intensity is $\ge$1/2 the maximum integrated
intensity. All the derived sizes are much larger than the telescope
resolution, and smaller than the map size.  The average FWHM linewidth
in the cores was determined by combining all the spectra in the
regions of the maps where emission was detected, and then fitting a
Gaussian profile to the resultant average spectrum. In a few cases
(e.g.\ Mon\,R2, S\,140 and Ceph\,A) this is likely to be affected by
strong outflows.  Note that for multiply-peaked or filamentary sources
the definitions of size we make here may not be appropriate; this
definition is most applicable to sources which show some degree of
circular symmetry.

We compare the linewidths of the \hbox{1--0} and \hbox{2--1}
transitions in figure~\ref{lw}.  The \xxco~2--1 linewidth is closely
correlated with the \xxco~1--0 linewidth with a best-fit slope close
to one. This is expected given the low optical depth of the \xxco\
transitions.  When the two linewidths do differ, the linewidth given
by \hbox{2--1} transition is in all but two cases smaller.  This is
likely an effect of the larger field covered with \xxco~1--0 -- the
inclusion of a larger region of the cloud should typically broaden the
average linewidth due to velocity shifts across each map.  In most
cases, however, the average linewidths are dominated by the stronger
line emission in the centers of each region, resulting in the strong
correlation seen in figure~\ref{lw}a.  The
\xco\ and \xxco~1--0 linewidths do not show the same degree of
correlation (figure~\ref{lw}b), with the \xco\ linewidths typically
being higher.  This likely due in part to the higher optical depth in
the \xco\ line.

We also calculate the virial mass, M$_{\rm V}$, from the tabulated
sizes and linewidths using the form of the virial theorem given in
\citet{rw00}:
\begin{equation}
\frac{M_{\rm V}}{M_\odot} = 
250 \left(\frac{\Delta v_{1/2}}{km\,s^{-1}}\right)^2
\left(\frac{R}{pc}\right).
\end{equation}

\noindent
The gas mass derived from the \xxco\ 1--0 data using the LTE analysis
and virial theorem are compared in figure \ref{virial}.  The majority
of sources show virial masses between one to two times larger than the
\xxco\ mass.  This difference will be reduced by the choice of the
virial constant; for clouds with $r^{-2}$ density profiles, the virial
masses we have calculated will be a factor of two too large
\citep{mac88}.  Furthermore, the virial theorem should only be
applicable to centrally condensed morhpologies, and should not apply
to the many regions showing extended or diffuse morphologies, or
exhibiting multiple peaks.  Nevertheless, the observation that most
virial masses are within a factor of two of the LTE mass suggests that
in most of the clouds, the gravitational energy dominates the kinetic
energy, and that most of the clouds are graviationally bound.

\citet{larson81} found that molecular clouds exhibit a well defined
relationship between cloud size and average linewidth.  In figure
\ref{size_lw}, we display the size-linewidth relation for the \xxco\
1--0 transition in our sample of regions. This transition should give
the most accurate relation because it is unaffected by optical depth
effects like the \hbox{\xco\ 1--0}, but encloses all of the emission,
unlike the \xxco\ 2--1.  We find considerable scatter compared to
Larson's relation; however, we cover only an order of magnitude in
core size, much smaller than the three orders of magnitude considered
by Larson.  The scatter may also be increased due to the method we
used to calculate the size of the region, which is primarily
applicable to clouds showing a circular symmetry.

Figure \ref{fir_lw}a shows the average \xxco\ 2--1 linewidth plotted
against FIR luminosity. There is a 3.7$\sigma$ correlation between
linewidth and luminosity. This is not due to a correlation with
distance, as is indicated in fig. \ref{fir_lw}b -- the significance of
the correlation between linewidth and distance is only 1.8$\sigma$.

\section{Summary}
\label{summ}

We have begun a program of multi-wavelength observations which, when
combined with existing infrared and optical data will yield detailed
information about both the stars and their surrounding gas in a large
sample of young stellar clusters within 1\,kpc of the sun.
Here we presented millimeter spectral line maps of a sample of 30 of
these cluster-forming regions in the \hbox{1--0} transitions of \xco\
and \xxco. Smaller regions surrounding 17 of these sources were also
mapped in the \xxco\ \hbox{2--1} transition.  Based on our \xco~1--0
maps and the location of the stellar or protostellar source we
proposed a sequence for the morphology of gas
surrounding the clusters. The youngest sources have centrally
condensed cores containing an embedded source, while the more evolved
sources have extended or diffuse molecular gas. Optically
visible Herbig Ae/Be stars associated with these sources are often
located in cavities, suggestive that these types of stars can be
responsible for the dispersal of molecular gas in cluster-forming
regions.

We determined cloud masses independently using the three observed
\xco\ and \xxco\ transitions, and compared the cloud properties derived from
each of the three tracers in order to better understand systematic
uncertainties in determining masses and linewidths.  We found good
consistency between the LTE and LVG methods, and between the masses
and linewidths derived from \xxco~1--0 and 2--1. Virial masses were
found to be within a factor of two of the LTE mass in most of the
regions, suggesting that the gravitational energy dominates the
kinetic energy in these regions, and that they are graviationally
bound.

Fits images of the integrated emission will be made available for
general use via the project website. Our final data set will include
sensitive ground based wide-field near-infrared imaging and
spectroscopy, millimeter spectral-line and continuum maps and
mid-infrared observations obtained with SIRTF. Once complete, it will
be made available to the community via a web-based database, providing
an invaluable resource for future star-formation studies.

\acknowledgments
FCRAO is supported by NSF grant AST\,01-00793. This work is based in
part on measurements made with the Heinrich-Hertz telescope, which is
operated by the Submillimeter Telescope Observatory on behalf of
Steward Observatory and the Max Planck Institut f\"ur
Radio\-astronomie. This work made use of the NASA/IPAC Infrared
Science Archive, which is operated by the Jet Propulsion Laboratory,
California Institute of Technology, under contract with the National
Aeronautics and Space Administration. We thank R. Gutermuth and
D. Peterson for assistance with the FCRAO observing.

\appendix
\section{Individual Sources}
\label{sources}

In this section we give a brief summary of previous observations of
each of the groups and clusters. For clarity we give a selection of
just the most recent references on each source.

\subsection{Class I}

\subsubsection{BD+40\degrees4124}
The Herbig Be star BD+40\degrees4124 is the most massive member of a
small group of young emission line stars \citep{hmss95}, and a large
number of highly embedded stars detected in the infrared by
\citet{palla95}. There is also a small molecular outflow in the
region, probably driven by the embedded source
\objectname{V1318S} \citep{palla95}.

\subsubsection{S\,131}      
S\,131, also called Ceph\,OB Cloud 37 is a bright-rimmed cloud
associated with an H{\sc ii} region \citep*{sugi91} and the Galactic
open cluster \objectname{IC\,1396}. A small cluster was observed in
this region by \citet{sugi95}, and several H$\alpha$ emission-line
stars were detected by \citet*{osp02}. Interferometric \xxco\ observations 
of this region were made by \citet{sugi97}.

\subsubsection{S\,140}      
The S\,140 region is a prototypical cometary globule, photoionised on
the south-west side by the B0 star \objectname{HD\,211880}. It
contains at least three bright infrared sources. A high-velocity
molecular outflow was detected in \co\ observations by
\citet{ssse84}. Previous smaller scale molecular line observations
have been made by \citet*{mwp93} who determined that the outflow was
driven by the source IRS\,1. 

\subsubsection{L\,1206}     
L\,1206 is a bright-rimmed globule associated with the H{\sc ii}
region S\,145. It has a molecular outflow, discovered in a survey by
\citet{sugi89}. 

\subsubsection{Cep\,A}      
The Cepheus\,A region is a molecular cloud containing a very active
star forming region indicated by a powerful molecular outflow
\citep*{rmh80}, more than 25 water masers \citep{torelles96} and
numerous thermal and non-thermal radio sources
\citep{hughes01,garay96}. 

\subsubsection{AFGL\,490}   
AFGL\,490 is a relatively isolated infrared source in the Cam\,OB1
complex \citep*{chp91}.  It is a young massive star-forming region,
containing a cluster of embedded sources \citep{hodapp94} and
associated with a molecular outflow \citep{ssse84}, and maser emission
\citep*{henkel86}.  

\subsubsection{GGD\,4}    
GGD\,4 contains a small cluster of young stars \citep{hodapp94} and
the driving source of a molecular outflow \citep{fukui89}.  

\subsubsection{CB\,34}    
CB34 is a relatively distant large globule containing a small group of
infrared sources \citep{ay95}, at least two high-velocity jets
\citep{ksgsd02} and several millimeter continuum sources thought to be
very young protostars \citep*{hws00}.  

\subsubsection{Mon\,R2}
This is an association of B1-B9 stars located in the Mon\,R2 giant
molecular cloud. It is a very active star-forming region, associated
with a powerful molecular outflow
\citep{bl83,tbww97}, a compact H{\sc ii} region \citep{wc89}, an infrared
cluster \citep{carp2000}, and OH masers \citep{mcb01}.  

\subsubsection{GGD\,12-15} 
GGD\,12--15 is an active star-forming region embedded in the Monoceros
molecular cloud. It is associated with a strong water maser, a
cometary compact H{\sc ii} region \citep{rod78,rod80,rod98}, several
millimeter-continuum sources and a bipolar CO outflow
\citep*{lhd90}.  

\subsubsection{NGC\,2264}
The NGC\,2264 region in northern Monoceros contains two well studied
star formation regions, IRS\,1 and IRS\,2.  Thirty IRAS sources, most
classified as Class I protostars \citep*{mly89} are present, as well as
$\sim$360 near-infrared sources \citep*{lyg93}. At least nine molecular 
outflow sources were identified in an unbiased survey by \cite{ml86}.

\subsection{Class II}

\subsubsection{IRAS\,20050}
The region IRAS\,20050 contains an embedded star and an
extremely-high-velocity (EHV) molecular jet and multipolar molecular
outflow, likely the superposition of outflows from several young stars
\citep*{bft95}. 

\subsubsection{S\,106}    
The S\,106 region contains a bipolar H{\sc ii} region and an
associated molecular cloud. Near-infrared observations by \citet{hr91}
revealed a cluster of 160 young stars in this region. A recent much
deeper near-infrared survey of the region suggests that the total
number of young stars could be much greater ($\sim$ 600)
\citep{oasa02}. CO observations of this region were made by \citet{ssksb02}.

\subsubsection{NGC\,7129} 

NGC\,7129 is a reflection nebula in the region of a young cluster,
containing three B-type stars (\objectname{BD+65\degrees1638},
{\objectname{BD+65\degrees1637} and LkH$\alpha$234) and several
embedded infrared sources. It is associated with Herbig-Haro objects
and several molecular outflows \citep*[][ and
refs. therein]{fms01}. Submillimeter continuum observations by
\cite{fms01} reveal several sources which they interpret as
prestellar. \citetalias{fuente02} have made higher-resolution \xco\
and \xxco~1--0 observations of the molecular gas in a small portion of
this region, finding a cavity surrounding BD+65\degrees1637, while
LkH$\alpha$234 is located at the peak of \xco\ emission.
Molecular line observations of the region were also made by
\cite{mtmm01}.

\subsubsection{S\,140-N}
The region S\,140-N contains a molecular outflow, discovered by
\citet{fuk86}. \citet{davis98} made a higher resolution \co~2--1
map of the outflow and determined that it is powered by the
intermediate-luminosity IRAS source
\objectname{22178+6317}. To the east of the molecular outflow source
are a sequence of Herbig-Haro objects (\objectname{HH\,251-254})
oriented in a northwest-southeast direction indicating a second
outflow \citep{eiroa93}.

\subsubsection{L\,1211}
L\,1211 is a dense core in the Cepheus cloud complex. It contains a
very young cluster, detected in millimeter continuum observations by
\citet{tmmb99}, and powers two molecular outflows.

\subsubsection{Cep\,C}
The Cepheus\,C region is an isolated core within the Cepheus\,OB3
molecular cloud \citep{sargent77}. The region contains a cluster of
infrared sources \citep{hodapp94} and is associated with water maser
emission \citep{ww86} and an outflow \citep{fukui89}.  

\subsubsection{S\,171}    
S\,171 is a bright-rimmed cloud associated with the Cepheus OB4
stellar association, an actively star-forming H{\sc ii} region, and
molecular cloud complex \citep{yf92}. The main heating and ionizing
source of this region is believed to be the star cluster,
\objectname{Be\,59} \citep{oosd02}. Several H$\alpha$ emission-line
stars were detected in this region by \citet{osp02}.

\subsubsection{NGC\,1333} 
NGC\,1333 is a reflection nebula associated with a region of recent,
extrememly active star formation in the Perseus molecular cloud. A
cluster of about 150 low- to intermediate-mass YSOs have been
identified in near-infrared images \citep*{asr94}. The region also
contains about 20 groups of Herbig-Haro objects, some with highly
collimated jets \citep*{bdr96}.  \cite*{rod99} found a total of 44
sources at centimeter wavelengths, most of which are associated or
believed to be associated with young stellar objects in the region.

\subsubsection{L\,1551}
L\,1551 is a compact molecular cloud located in the southern part of
the Taurus region. It shows abundant signs of ongoing star-formation,
and contains the infrared source IRS\,5, probably the most
well-studied low-mass young stellar object in the galaxy. IRS\,5
powers a powerful well-collimated bipolar molecular outflow
\citep[e.g.][]{mss88}, and is associated with several Herbig-Haro objects.
Also in the region are several other overlapping molecular outflows
and a number of T-Tauri stars (including the well-known sources
HL\,Tau and XZ\,Tau).

\subsubsection{VY\,Mon}   
VY\,Mon is a highly-reddened Herbig Be star in the
\objectname{Mon\,OB1} region.  It is located 85$''$ south of the
reflection nebula IC 446.  

\subsection{Class III}

\subsubsection{MWC\,297}
  
MWC\,297 is an extremely reddened Herbig Be star \citep{drew97} seen
in projection against the H{\sc ii} region \objectname{S\,62},
although its relationship to the H{\sc ii} region is not
clear. \citetalias{fuente02} have made higher-resolution \xco\ and
\xxco~1--0 observations of a small region surrounding this
source. Their observations show evidence for a cavity surrounding the
star, within a more diffuse molecular gas environment.

\subsubsection{VV\,Ser}
VV\,Ser was determined to be a Herbig Ae star by
\citet{testi98}. A small region surrounding this star  was mapped 
in \xco\ and \xxco~1--0 at higher
resolution by
\citetalias{fuente02}. They detected very litle molecular gas in the vicinity
of this star, and interpreted that it was located in a cavity.

\subsubsection{HD\,200775}
HD\,200775 is a B3 star located at the northern edge of an elongated
molecular cloud. It is the illuminating star of the well-known
reflection nebula \objectname{NGC\,7023} and is associated with a
bipolar outflow \citep{watt86} and a cluster of H$\alpha$
emission-line stars \citep{umhd01}.
The region has been previously mapped in
\xco~1--0 by \citet{hd200} and \citetalias{fuente02}. These observations
show that the star is located within a biconical cavity, that has
probably been excavated by a bipolar outflow. However \citet{hd200}
found no evidence for current high velocity gas within the lobes of
the cavity.

\subsubsection{L\,988-e}
The L\,988-e cloud contains a cluster of bright infrared sources
\citep{hodapp94} coinciding with the position
of a molecular outflow \citep{clark86}.  

\subsubsection{IC\,5146}
IC\,5146 is a young stellar cluster, associated with a reflection
nebula illuminated by the B0\,V star
\objectname{BD+46\degrees3474}. Also embedded in the same cloud is the
HAeBe variable \objectname{BD+46\degrees3471} and $\sim$100 H$\alpha$
emission-line stars \citep{herbig02}.

\subsubsection{HD\,216629}  
HD\,216629 is a Be star in the Cepheus cloud with a close companion
\citep*{pirzkal97} and surrounded by a cluster of infrared sources
\citep{testi98}. \citet{fuente98,fuente02} have made
\xco~1--0, \xxco~1--0 and 
CO~2--1 observations of a 1\,pc region surrounding this star. They
detected very little molecular gas, as traced by \xco\ close to the
star. They describe the gas morphology in this source as a cavity.

\subsubsection{IC\,348}
The young cluster IC\,348 is located in the Perseus molecular
cloud. As well as $\sim$40 optically visible stars, including the
B5\,V star BD+31\degrees643 (associated with a well-known reflection
nebula), and 16 H$\alpha$ emission-line stars, it contains almost 400
embedded stars, detected in the near-infrared by
\citet{ll95}. High-resolution \xco\ and \xxco\ observations were made over
a small portion of this region by \citet{bgk87}.

\subsubsection{LkH$\alpha$101} 
LkH$\alpha$101 is a bright Herbig Ae/Be star associated with a
powerful ionised stellar wind, extended H{\sc ii} region and
reflection nebulosity \citep{bsk91}. \citet{ut87} associated this star
with an extension of the Perseus cloud, and the region containing
XY\,Per. Using an analysis of the reddening of Hipparcos stars
A. Wilson (2002, personal communication) determined a distance of 280\,pc to
this complex. We therefore adopt this distance, rather than the value
of 800\,pc quoted in the literature.

\subsubsection{XY\,Per}     
XY\,Per is a Herbig Be star surrounded by a small cluster of infrared
sources \citep{testi97}. It is located in the same molecular cloud as
LkH$\alpha$101 \citep{ut87}.

\newpage
\bibliographystyle{apj}
\bibliography{refs}

%---------------------------------------------------------------------------
%%FIGURES
\newpage
\onecolumn
\begin{figure}
\plottwo{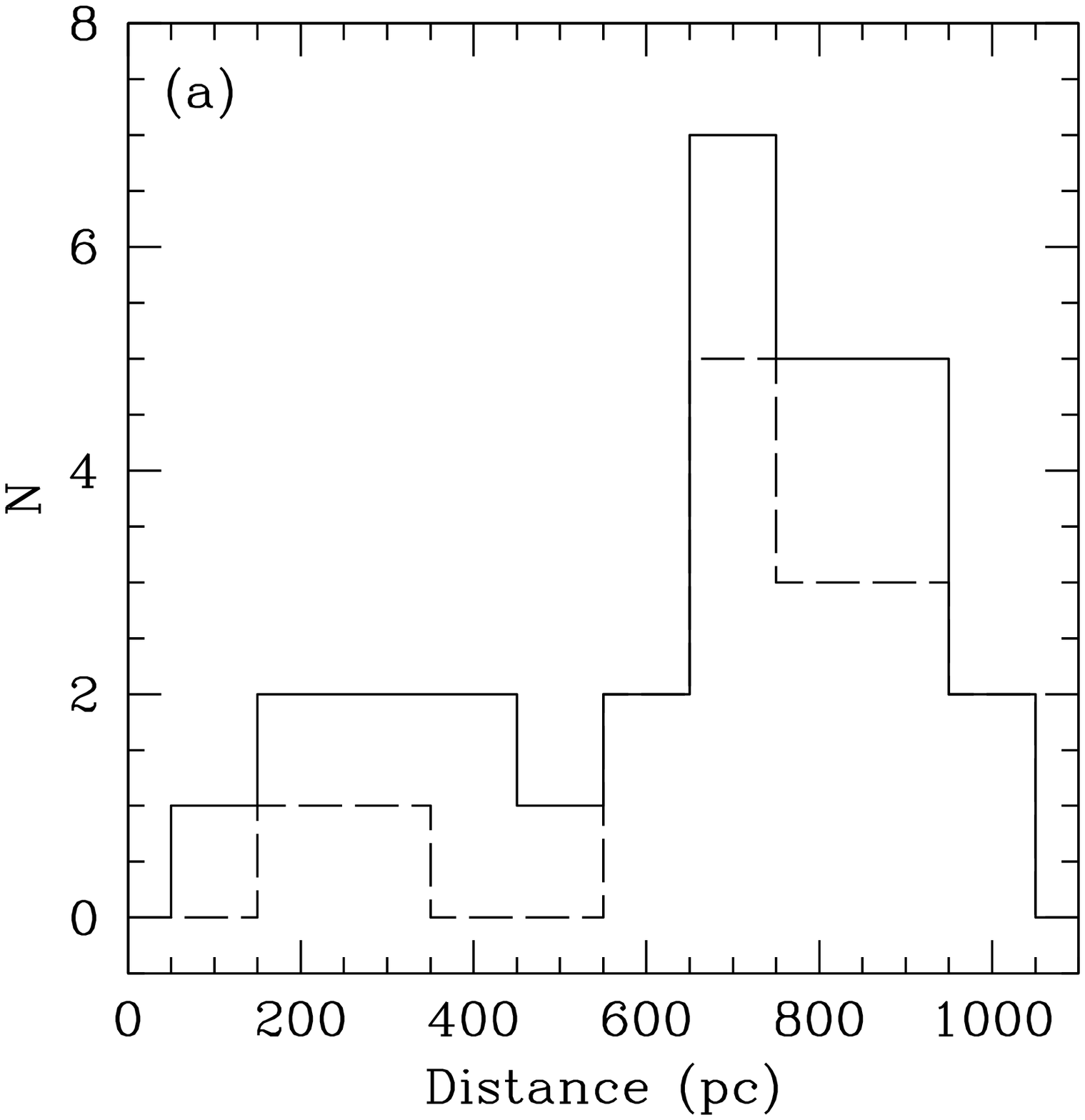}{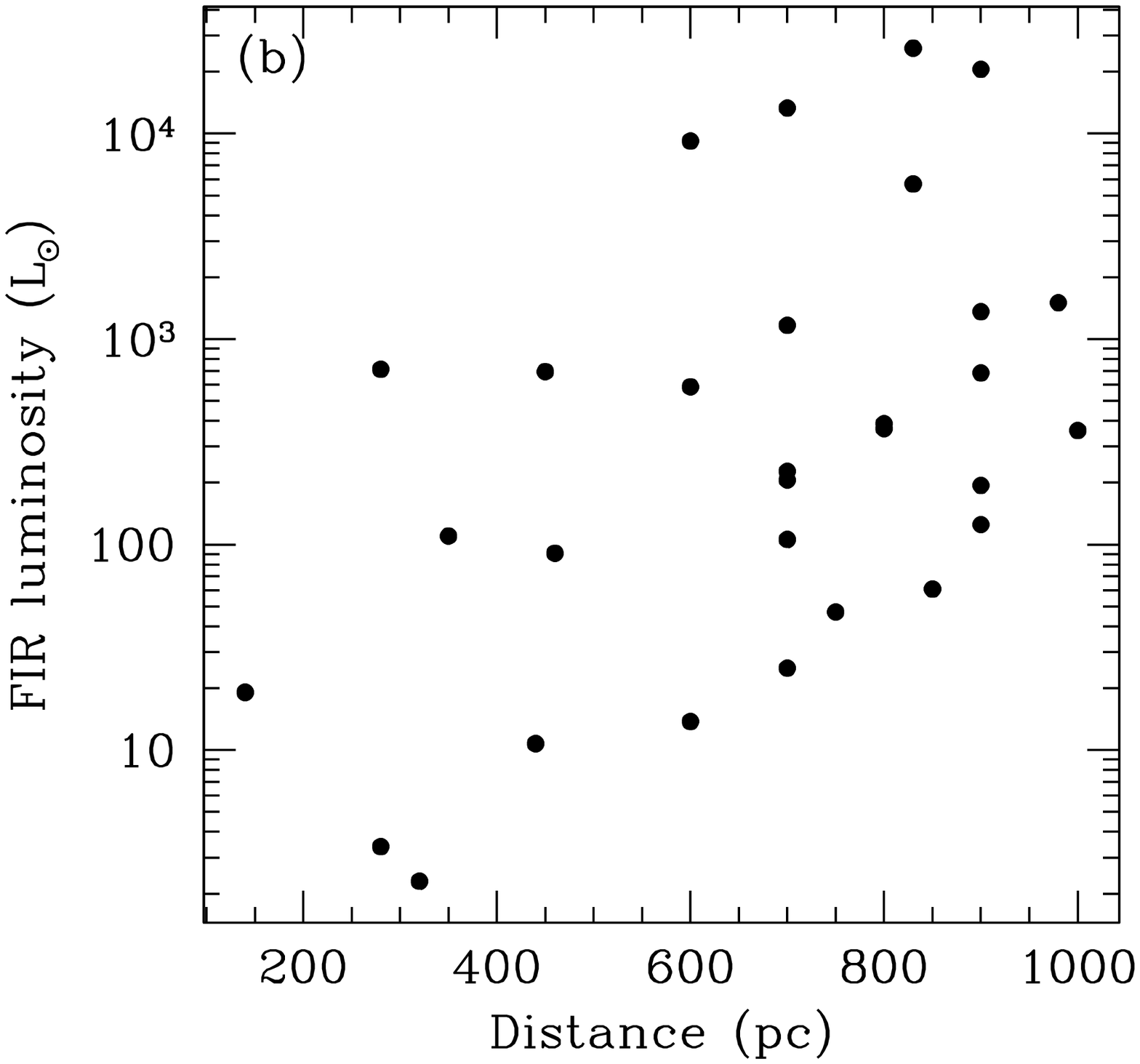}
\caption{(a) Histogram of number of sources with distance. The solid line indicates
the full sample, and the dashed line those sources for which all three
lines were observed.  (b) FIR luminosity vs. distance for all the
sources in the sample.
\label{dist}}
\end{figure}

%%---------------------------------------------------------------------------
%% FCRAO maps

\onecolumn
%%\clearpage
\begin{figure}
\epsscale{0.68}
\plotone{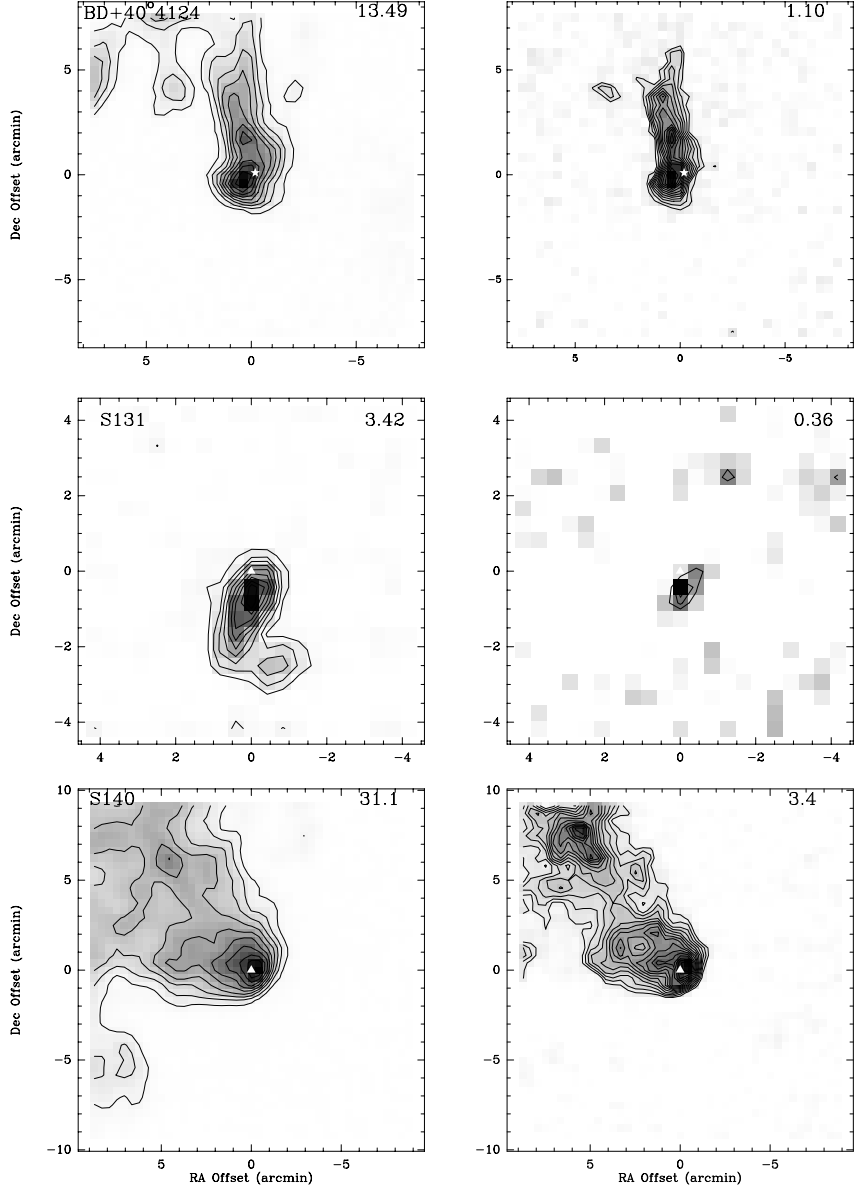}
\caption{Integrated intensity maps of BD+40\degrees4124 (top) and 
S\,131 (middle) 
and S\,140 (bottom), all classified as developmental Class I. The value
of the maximum integrated intensity in K\,\kms\ is given in the top
right corner of each panel.  Left panels show \xco\ 1--0 integrated
emission, with contours at 10\% intervals from 10\% to 100\% of the
maximum integrated intensity. Right panels show integrated \xxco\ 1--0
emission. Base contour is 3$\sigma$, contour interval is 1$\sigma$,
except for S\,131 where the base contour is at 2$\sigma$ and contour
intervals at 0.5$\sigma$.  (0,0) positions are as given in table
\ref{sourcetab}. Triangles indicate the position of an IRAS 
or near-infrared (embedded)
source 
and
stars indicate the position of an optically visible Herbig Ae/Be star.
\label{cont0}}
\end{figure}

%%\clearpage
\begin{figure}
\epsscale{0.7}
\plotone{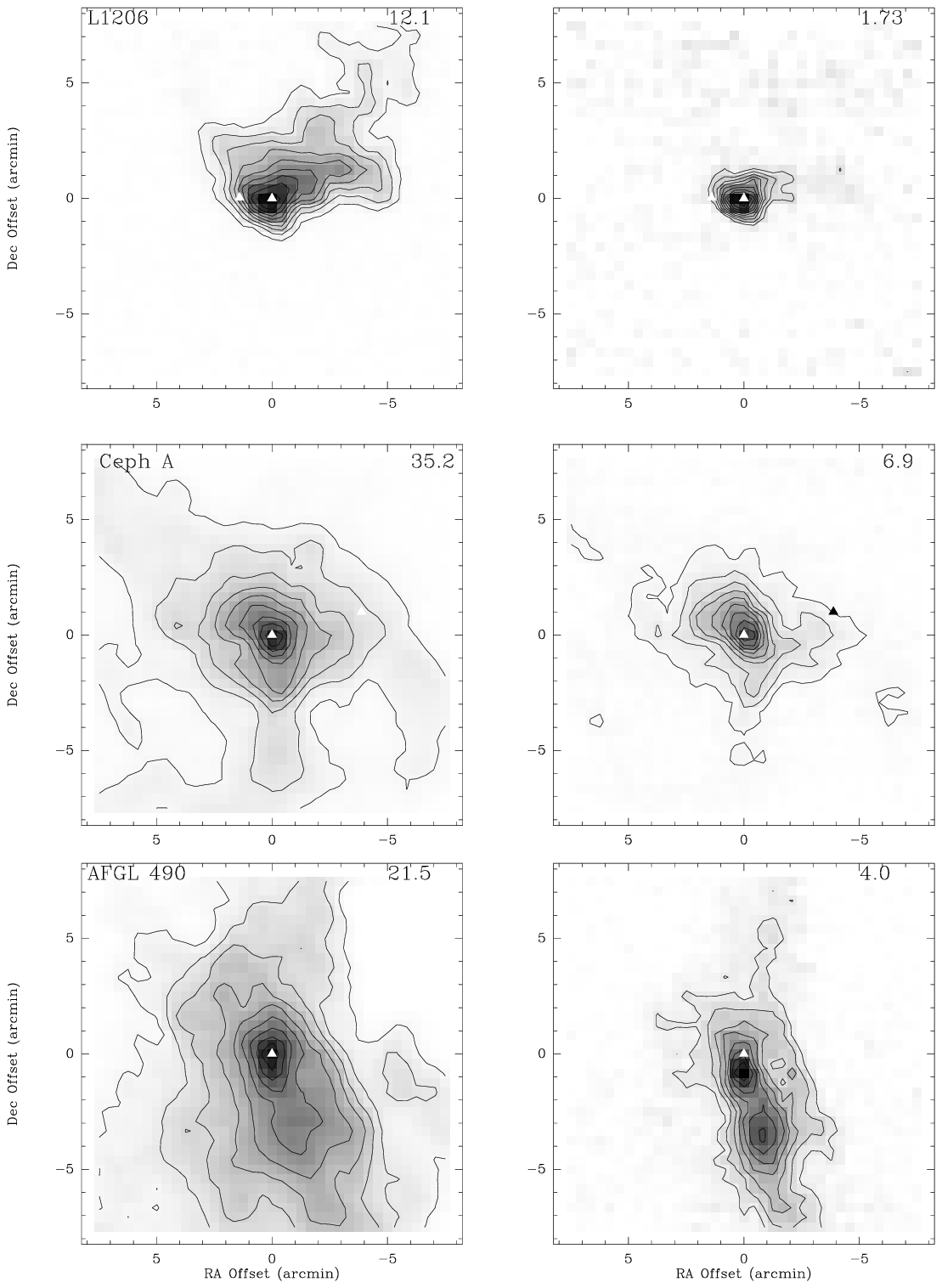}
\caption{Integrated intensity maps of developmental Class I sources
L\,1206 (top), Cep\,A (middle) and AFGL\,490 (bottom). 
The value of the maximum integrated intensity in K\,\kms\ is given in
the top right corner of each panel.  Left panels show \xco\ 1--0
integrated emission, with contours at 10\% intervals from 10\% to
100\% of the maximum integrated intensity. Right panels show
integrated \xxco\ 1--0 emission. Base contour is 3$\sigma$, contour
interval is 10\%, except for L\,1206 where contour interval is
1$\sigma$. (0,0) positions are as given in table
\ref{sourcetab}. Symbols as in figure \ref{cont0}.
\label{cont1}}
\end{figure}

%\clearpage
\begin{figure}
\epsscale{0.7}
\plotone{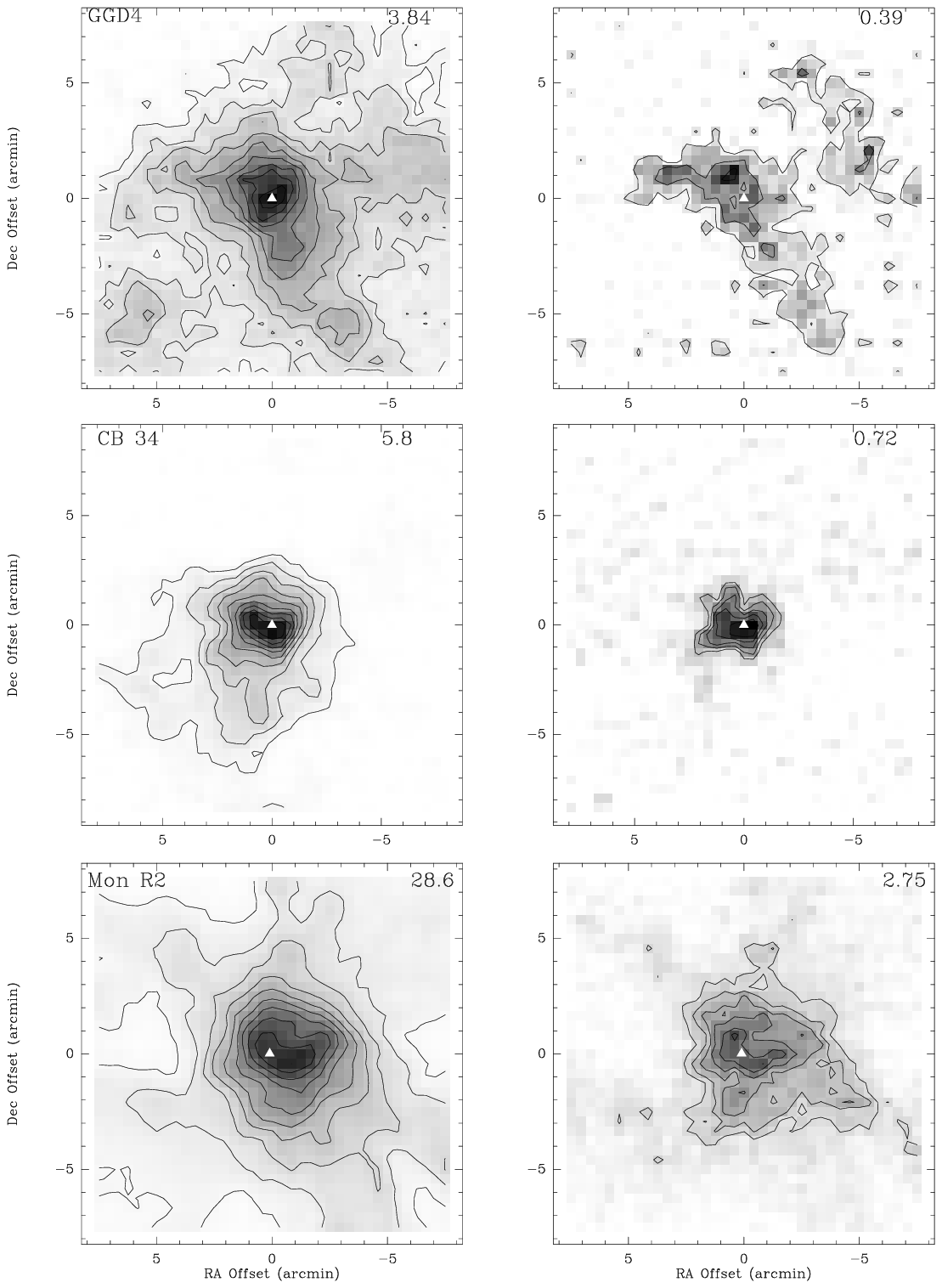}
\caption{Integrated intensity maps of GGD\,4 (top), CB\,34 (middle) and Mon\,R2 (bottom). 
The value of the maximum integrated intensity in K\,\kms\ is given in
the top right corner of each panel.  Left panels show \xco\ 1--0
integrated emission, with contours at 10\% intervals from 10\% to
100\% of the maximum integrated intensity. Right panels show
integrated \xxco\ 1--0 emission. Base contour is 3$\sigma$, contour
interval is 1$\sigma$, except for GGD\,4 where the base contour is at
1$\sigma$. (0,0) positions are as given in table
\ref{sourcetab}. Symbols as in figure \ref{cont0}.
\label{cont2}}
\end{figure}

%\clearpage
\begin{figure}
\epsscale{0.7}
\plotone{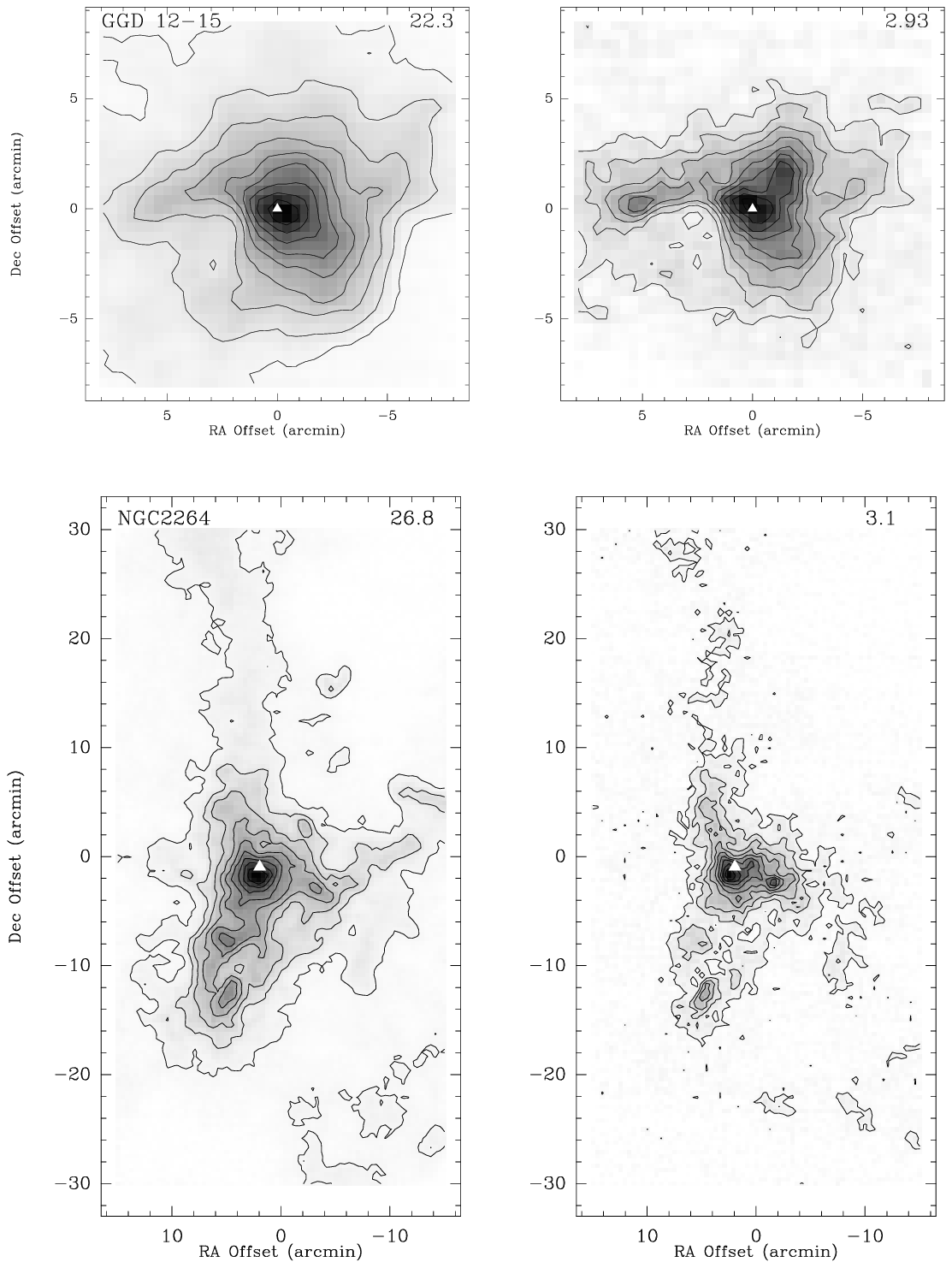}
\caption{Integrated intensity maps of 
Class I sources GGD\,12-15 (top) and NGC\,2264 (bottom).  The value of
the maximum integrated intensity in K\,\kms\ is given in the top right
corner of each panel.  Left panel shows \xco\ 1--0 integrated
emission, with contours at 10\% intervals from 10\% to 100\% of the
maximum integrated intensity. Right panel shows integrated \xxco\ 1--0
emission. Base contour is 3$\sigma$ (10\% for NGC\,2264), contour
interval is 1$\sigma$ (10\% for NGC\,2264). (0,0) positions are as
given in table
\ref{sourcetab}. Symbols as in figure \ref{cont0}.
\label{cont3}}
\end{figure}

\begin{figure}
\epsscale{0.69}
\plotone{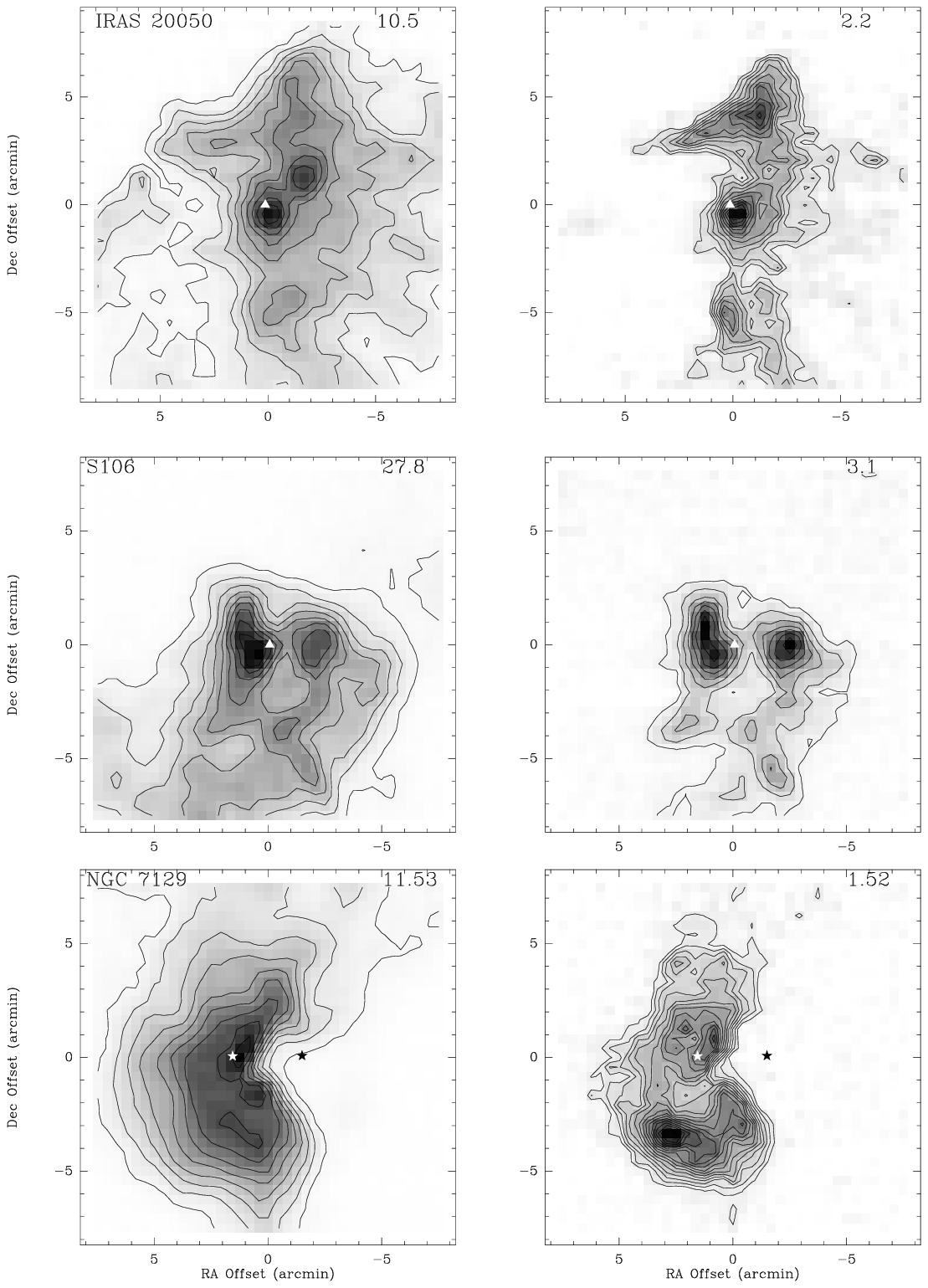}
\caption{Integrated intensity maps of Class II sources IRAS\,20050 (top), 
S\,106 (middle) and NGC\,7129 (bottom). 
The value of the maximum integrated intensity in K\,\kms\ is given in
the top right corner of each panel.  Left panels show \xco\ 1--0
integrated emission, with contours at 10\% intervals from 10\% to
100\% of the maximum integrated intensity. Right panels show
integrated \xxco\ 1--0 emission. Base contour is 3$\sigma$, contour
interval is 1$\sigma$.
(0,0) positions
are as given in table \ref{sourcetab}. Symbols as in figure
\ref{cont0}.
\label{cont6}}
\end{figure}

%\clearpage
\begin{figure}
\epsscale{0.7}
\plotone{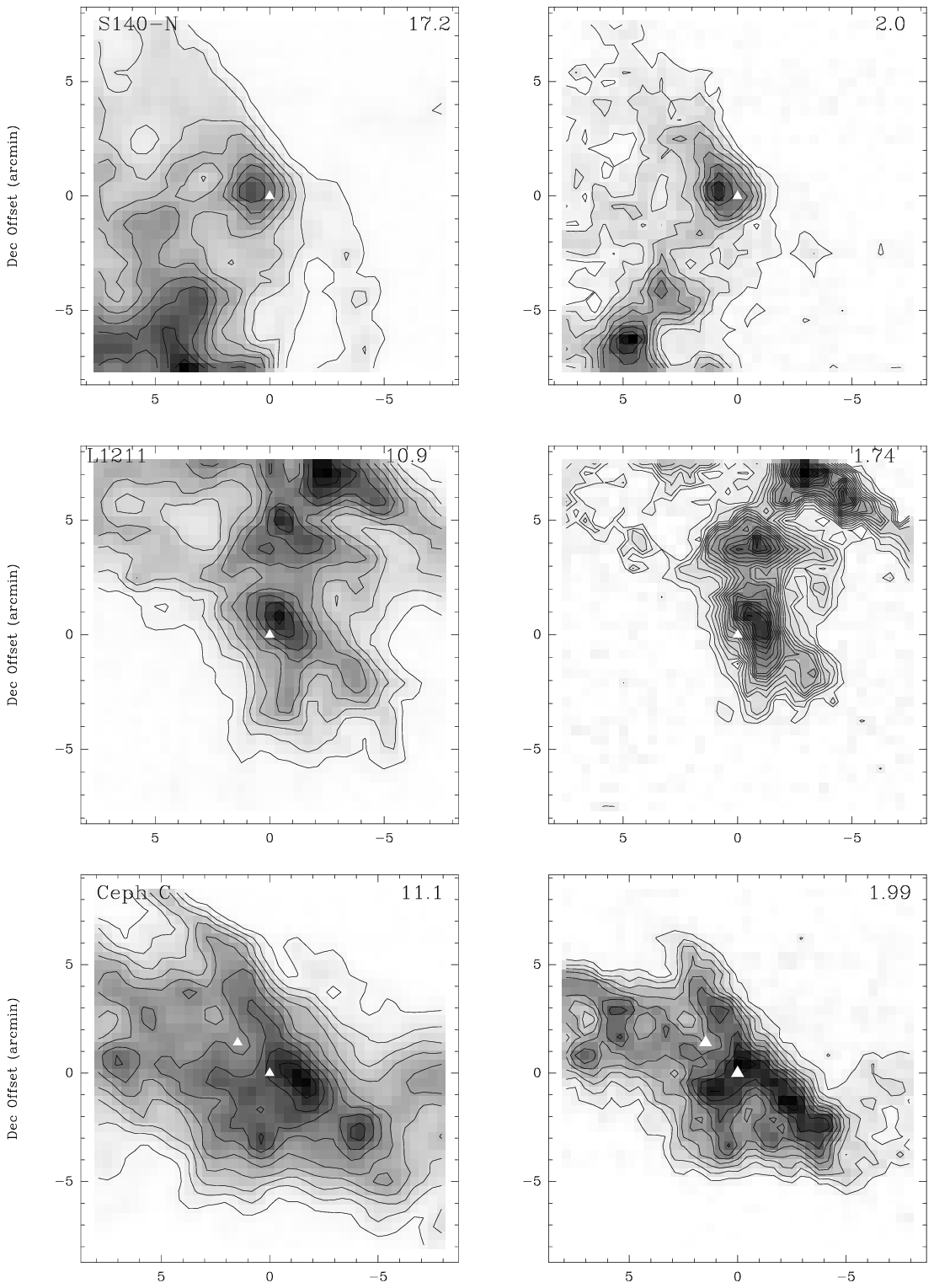}
\caption{Integrated intensity maps of Class II sources S\,140-N (top), 
L\,1211 (middle) and Ceph\,C (bottom). 
The value of the maximum integrated intensity in K\,\kms\ is given in
the top right corner of each panel.  Left panels show \xco\ 1--0
integrated emission, with contours at 10\% intervals from 10\% to
100\% of the maximum integrated intensity. Right panels show
integrated \xxco\ 1--0 emission. Base contour is 3$\sigma$ (10\% for S140-N), contour
interval is 1$\sigma$ (10\% for S140-N). 
(0,0) positions are as given in table
\ref{sourcetab}. Symbols as in figure \ref{cont0}.
\label{cont4}}
\end{figure}

%\clearpage
\begin{figure}
\epsscale{0.9}
\plotone{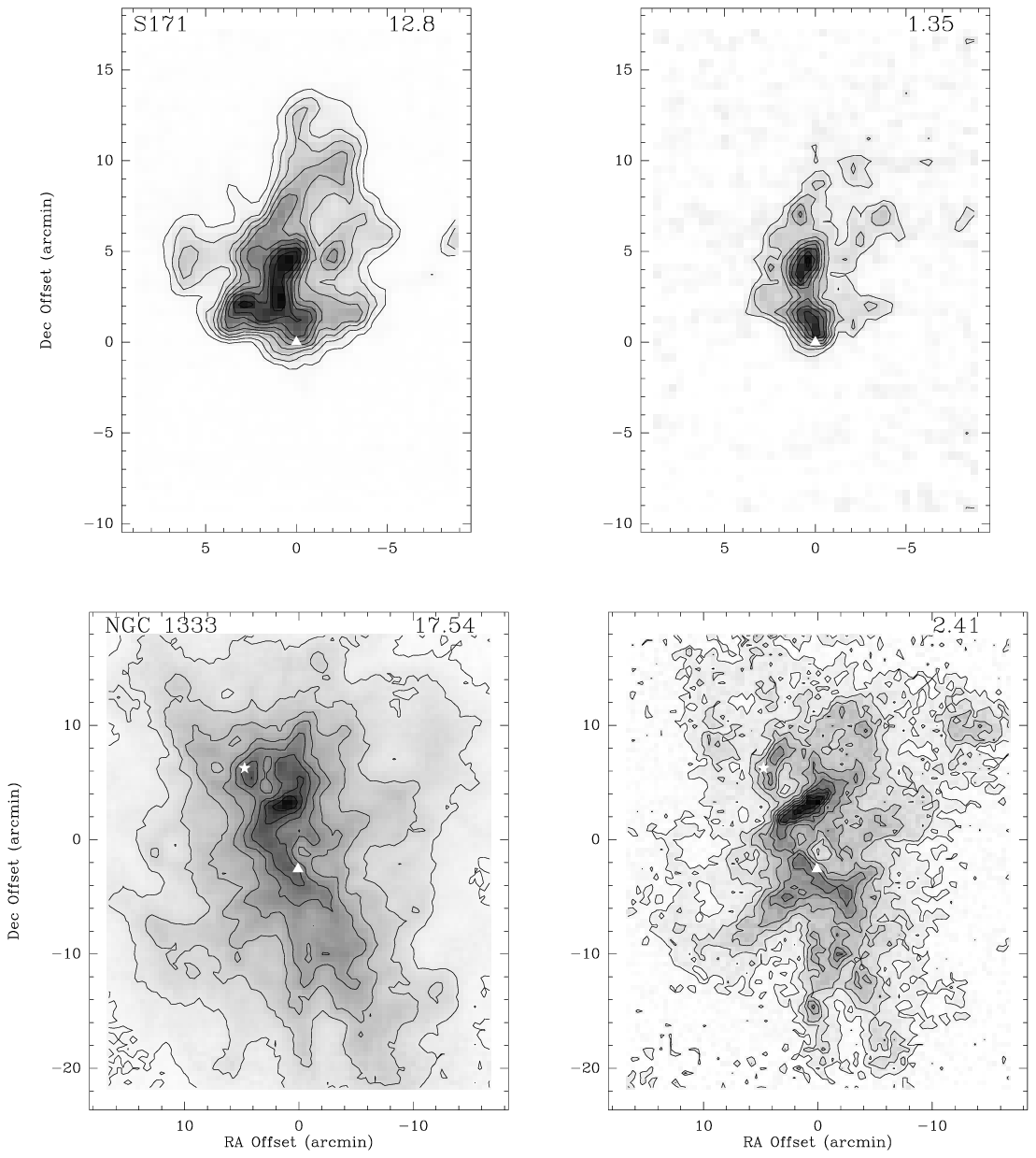}
\caption{Integrated intensity maps of Class II sources S\,171 (top) and NGC\,1333 (bottom). 
The value of the maximum
integrated intensity in K\,\kms\ is given in the top right corner of
each panel.  Left panels show \xco\ 1--0 integrated emission, with
contours at 10\% intervals from 10\% to 100\% of the maximum
integrated intensity. Right panels show integrated \xxco\ 1--0
emission. Base contour is 3$\sigma$, contour interval is 1$\sigma$. 
(0,0) positions are as given in table
\ref{sourcetab}. Symbols as in figure \ref{cont0}.
\label{cont5}}
\end{figure}

\begin{figure}
\epsscale{}
\plotone{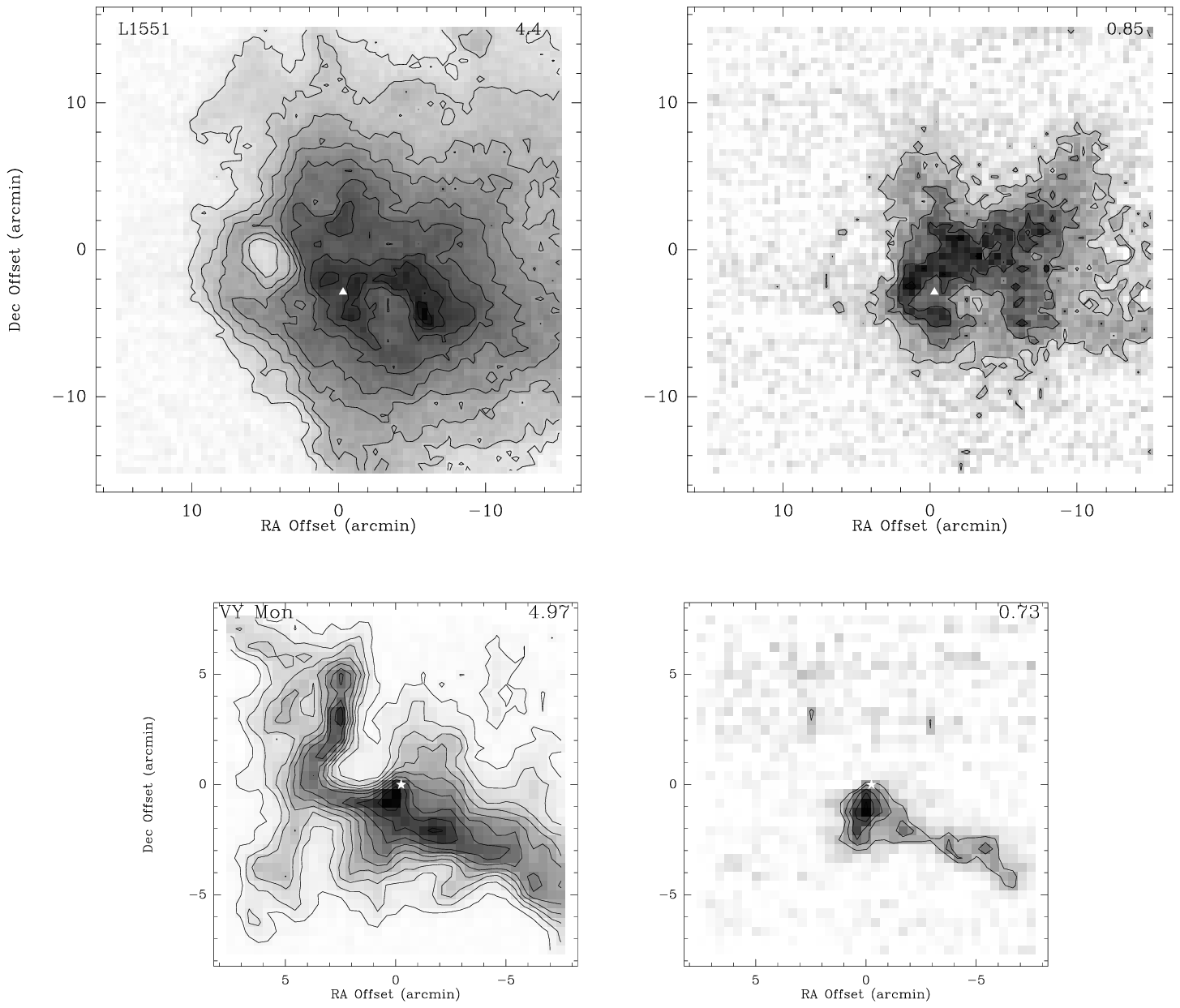}
\caption{Integrated intensity maps of Class II sources L\,1551 (top), 
and VY\,Mon (bottom). The value of the maximum
integrated intensity in K\,\kms\ is given in the top right corner of
each panel.  Left panels show \xco\ 1--0 integrated emission, with
contours at 10\% intervals from 10\% to 100\%
of the maximum integrated intensity. Right panels show integrated
\xxco\ 1--0 emission. Base contour is 3$\sigma$ (1$\sigma$ for L1551), contour interval is
1$\sigma$. (0,0) positions are as given in table
\ref{sourcetab}. Symbols as in figure \ref{cont0}.
\label{cont7}}
\end{figure}

\begin{figure}
\epsscale{0.7}
\plotone{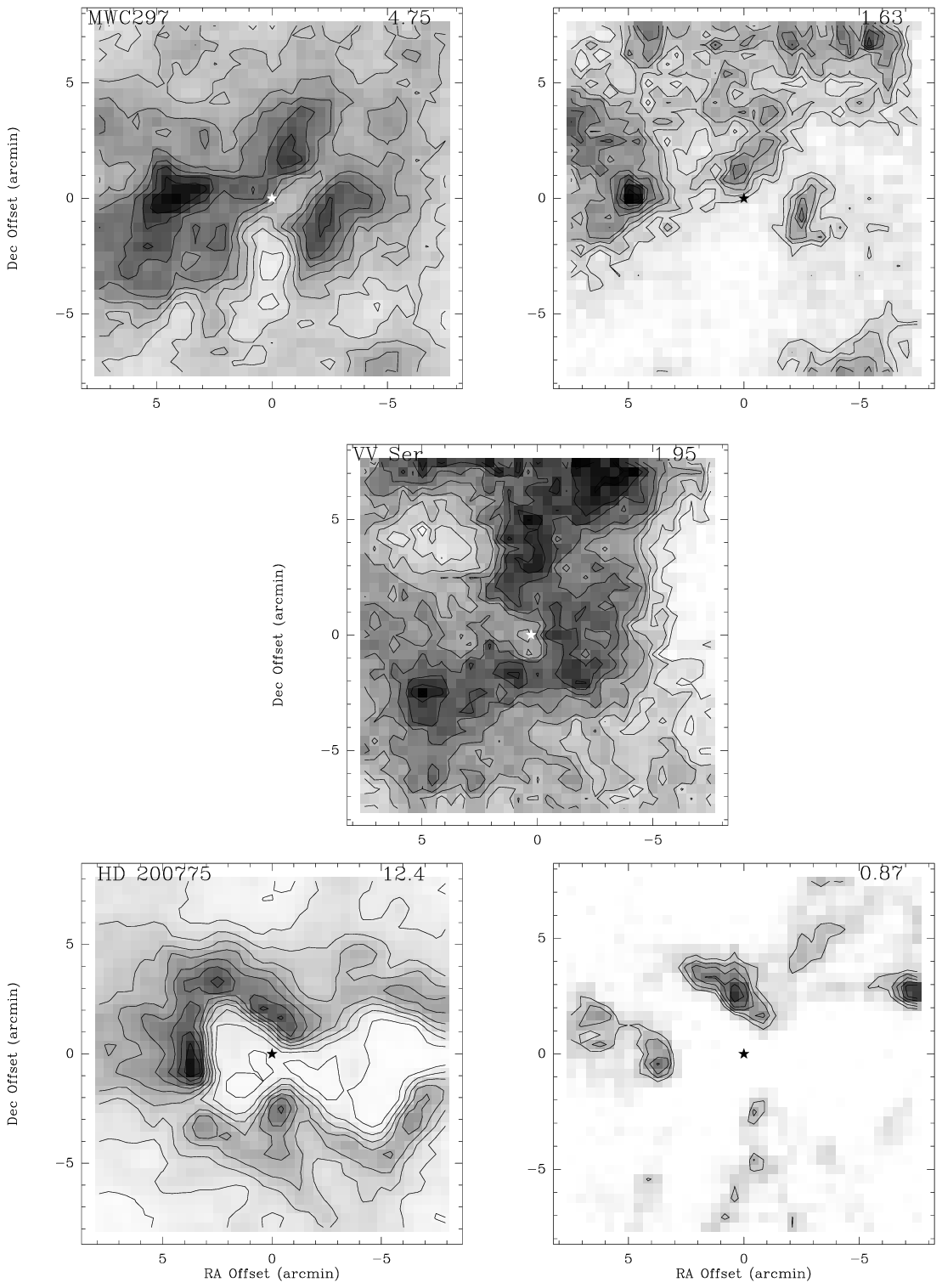}
\caption{Integrated intensity maps of Class III sources
MWC\,297 (top), VV\,Ser (middle) and HD\,200775 (bottom).
The value of the maximum
integrated intensity in K\,\kms\ is given in the top right corner of
each panel.  Left panels show \xco\ 1--0 integrated emission, with
contours at 10\% intervals from 10\% to 100\%
of the maximum integrated intensity. Right panels show integrated
\xxco\ 1--0 emission. Base contour is 3$\sigma$, 
contour interval is 1$\sigma$. \xxco\ emission was too weak to map in
VV\,Ser.  (0,0) positions are as given in table
\ref{sourcetab}. Symbols as in figure \ref{cont0}.
\label{cont8}}
\end{figure}
\clearpage 

\begin{figure}
%\epsscale{0.7}
\epsscale{}
\plotone{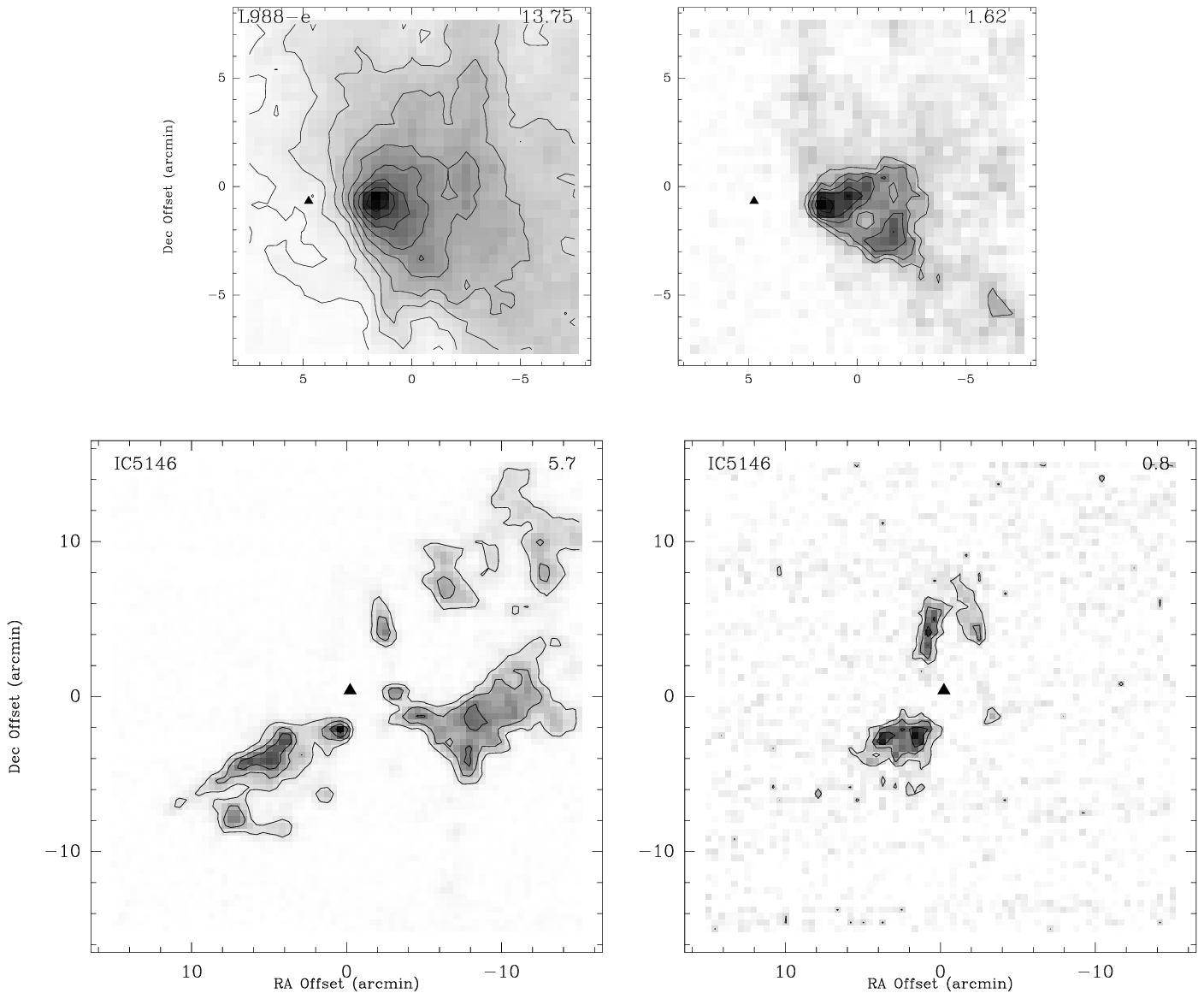}
\caption{Integrated intensity maps of Class III sources
L988-e (top) and IC\,5146 (bottom).
The value of the maximum
integrated intensity in K\,\kms\ is given in the top right corner of
each panel.  Left panels show \xco\ 1--0 integrated emission, with
contours at 10\% intervals from 10\% to 100\%
of the maximum integrated intensity. Right panels show integrated
\xxco\ 1--0 emission. Base contour is 3$\sigma$ (1$\sigma$ for IC\,5146), 
contour interval is 1$\sigma$. (0,0) positions are as given in table
\ref{sourcetab}. Symbols as in figure \ref{cont0}.
\label{cont9}}
\end{figure}
\clearpage

\begin{figure}
\epsscale{}
\plotone{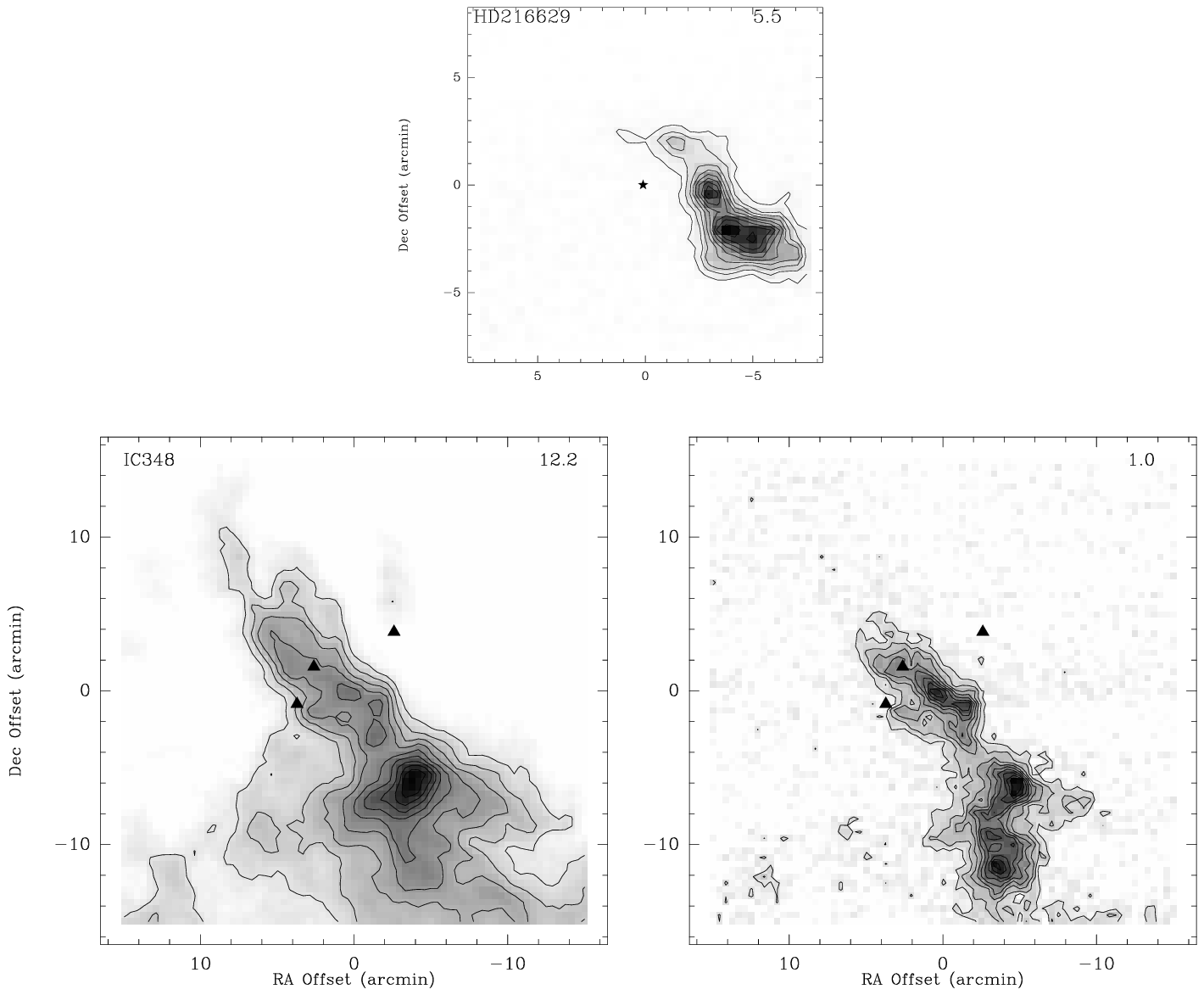}
\caption{Integrated intensity maps of Class III sources
HC\,216629 (top) and IC\,348 (bottom).
The value of the maximum
integrated intensity in K\,\kms\ is given in the top right corner of
each panel.  Left panels show \xco\ 1--0 integrated emission, with
contours at 10\% intervals from 10\% to 100\%
of the maximum integrated intensity. Right panels show integrated
\xxco\ 1--0 emission. Base contour is 3$\sigma$, 
contour interval is 1$\sigma$. (0,0) positions are
as given in table
\ref{sourcetab}. Symbols as in figure \ref{cont0}.
\label{cont10}}
\end{figure}
\clearpage 

\begin{figure}
\epsscale{0.7}
%\epsscale{}
\plotone{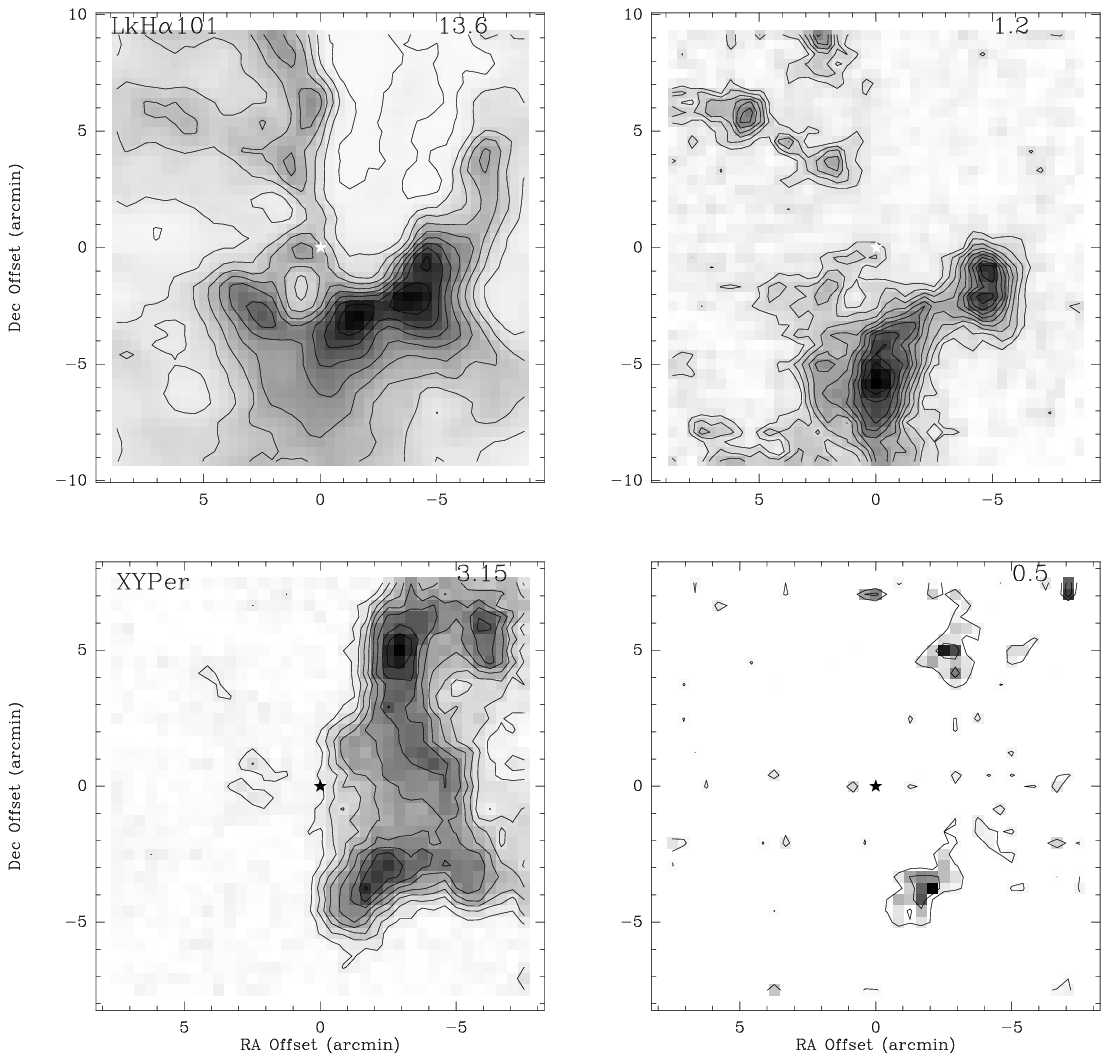}
\caption{Integrated intensity maps of Class III sources
LkH$\alpha$101 (top) and XY\,Per (bottom).
The value of the maximum
integrated intensity in K\,\kms\ is given in the top right corner of
each panel.  Left panels show \xco\ 1--0 integrated emission, with
contours at 10\% intervals from 10\% (3$\sigma$ for XY\,Per) to 100\%
of the maximum integrated intensity. Right panels show integrated
\xxco\ 1--0 emission. Base contour is 3$\sigma$ (1$\sigma$ for XY\,Per), 
contour interval is 1$\sigma$. (0,0) positions are
as given in table
\ref{sourcetab}. Symbols as in figure \ref{cont0}.
\label{cont11}}
\end{figure}
\clearpage 

%%------------------------------------------------------------------------------------------
%% SMTO maps

\twocolumn
\begin{figure}
\epsscale{}
\plotone{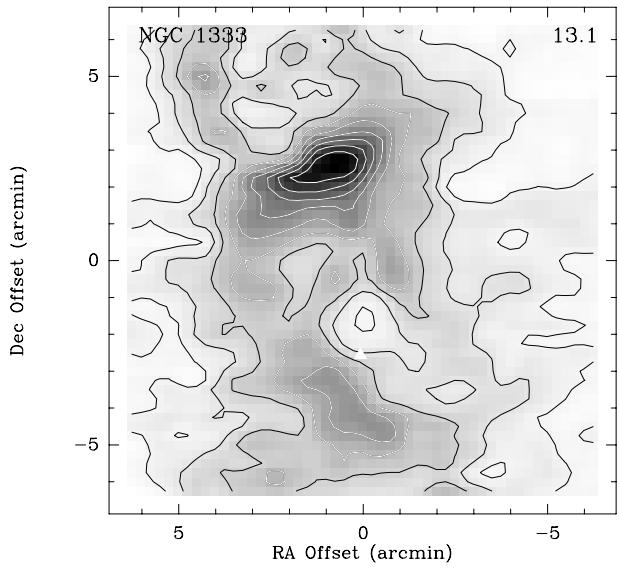}
\caption{\xxco\ 2--1 integrated intensity map of NGC\,1333.
Base contour is at 10\%, contour interval is 10\% of the maximum
integrated intensity, as given in the top right corner of the
panel. (0,0) positions are as given in table \ref{sourcetab}. Symbols
are as in figure \ref{cont0}.
\label{smto0}}
\end{figure}

\onecolumn
\begin{figure}
\epsscale{0.8}
\plotone{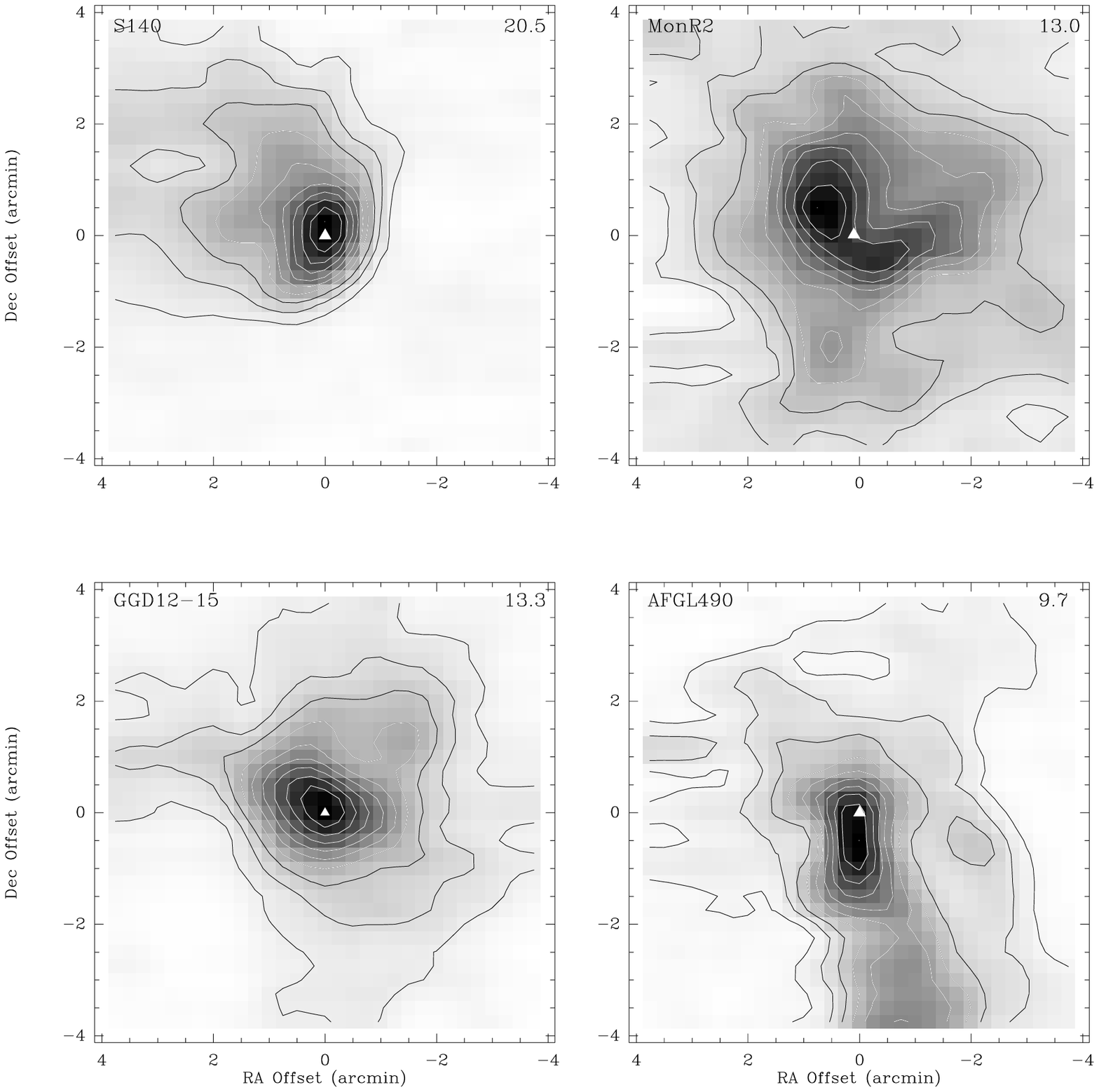}
\caption{\xxco\ 2--1 integrated intensity maps of S140, Mon\,R2, GGD\,12-15 and AFGL\,490.
Base contour is at 10\%, contour interval is 10\% of the maximum
integrated intensity, as given in the top right corner of each
panel. (0,0) positions are as given in table \ref{sourcetab}. Symbols
are as in figures \ref{cont0}.
\label{smto1}}
\end{figure}

\begin{figure}
\epsscale{0.7}
\plotone{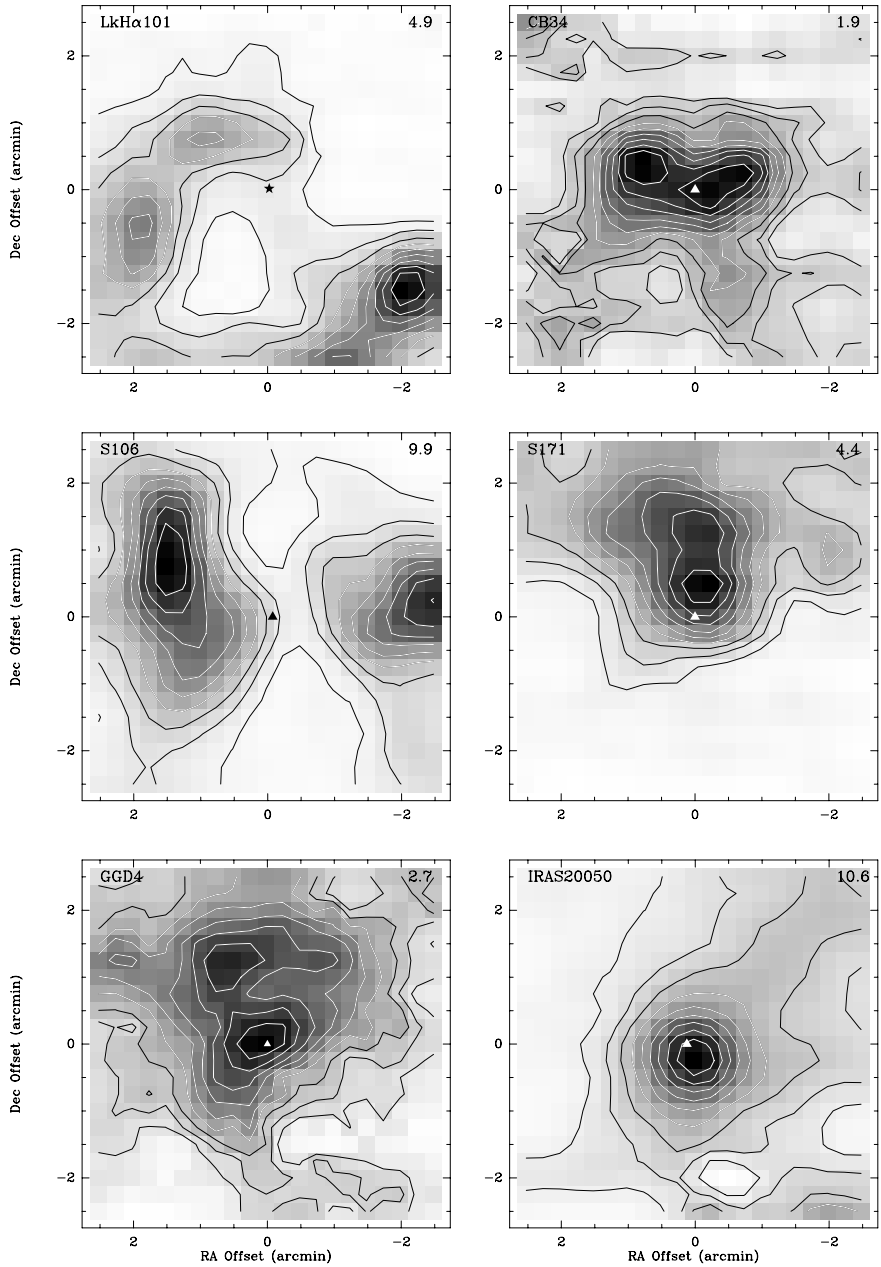}
\caption{\xxco\ 2--1 integrated intensity maps of LkH$\alpha$101, CB34, S106, S171, GGD4, and IRAS\,20050.
Base contour is at 10\%, contour interval is 10\% of the maximum
integrated intensity, as given in the top right corner of each
panel. (0,0) positions are as given in table \ref{sourcetab}. Symbols
are as in figures \ref{cont0}.
\label{smto2}}
\end{figure}

\begin{figure}
\epsscale{0.7}
\plotone{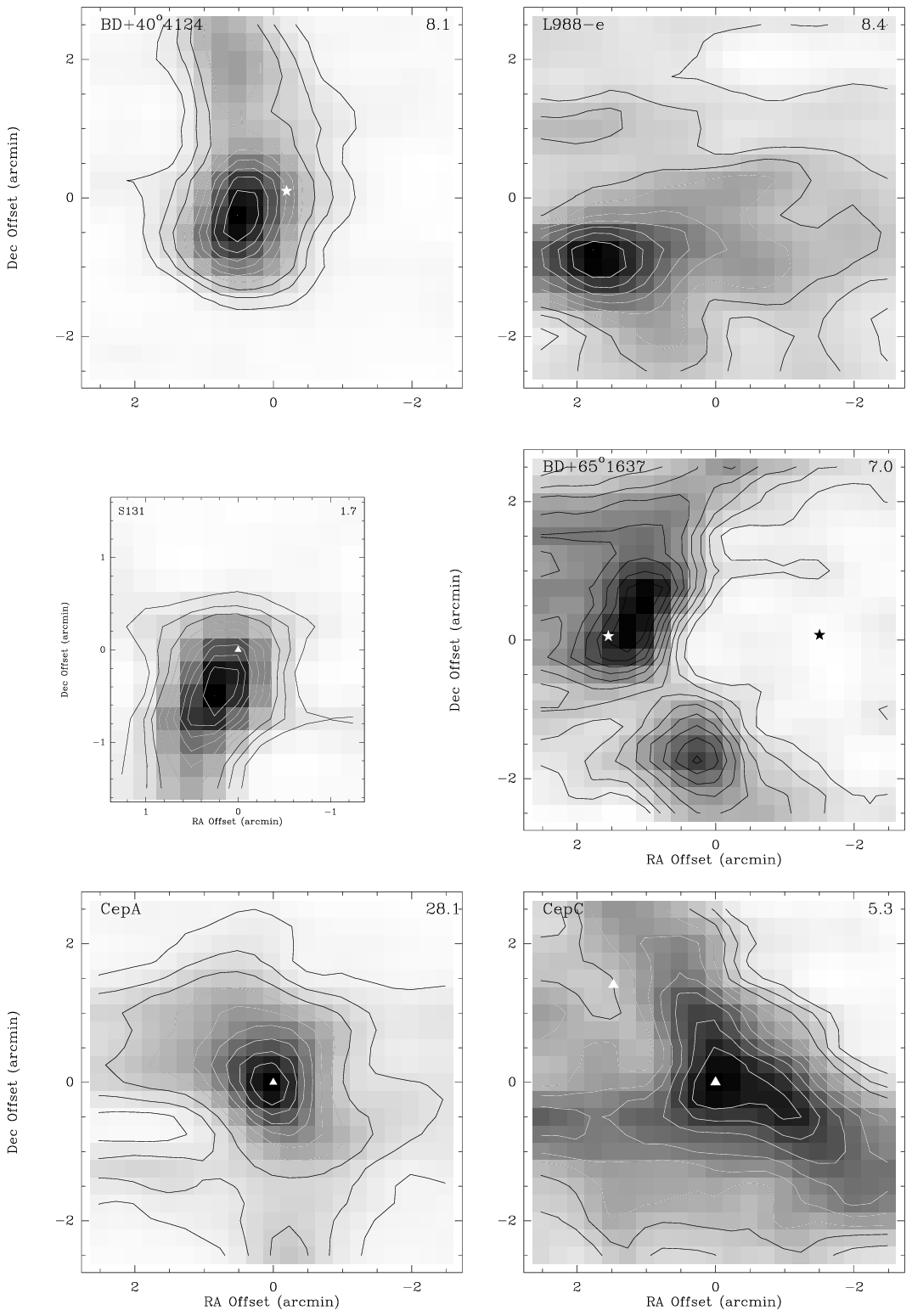}
\caption{\xxco\ 2--1 integrated intensity maps of BD+40\degrees4124, L988-e, S131, NGC\,7129, Cep\,A and Cep\,C.
Base contour is at 10\%, contour interval is 10\% of the maximum
integrated intensity, as given in the top right corner of each
panel. (0,0) positions are as given in table \ref{sourcetab}. Symbols
are as in figures \ref{cont0}.
\label{smto3}}
\end{figure}

%\onecolumn
\begin{figure}
\epsscale{0.9}
\plotone{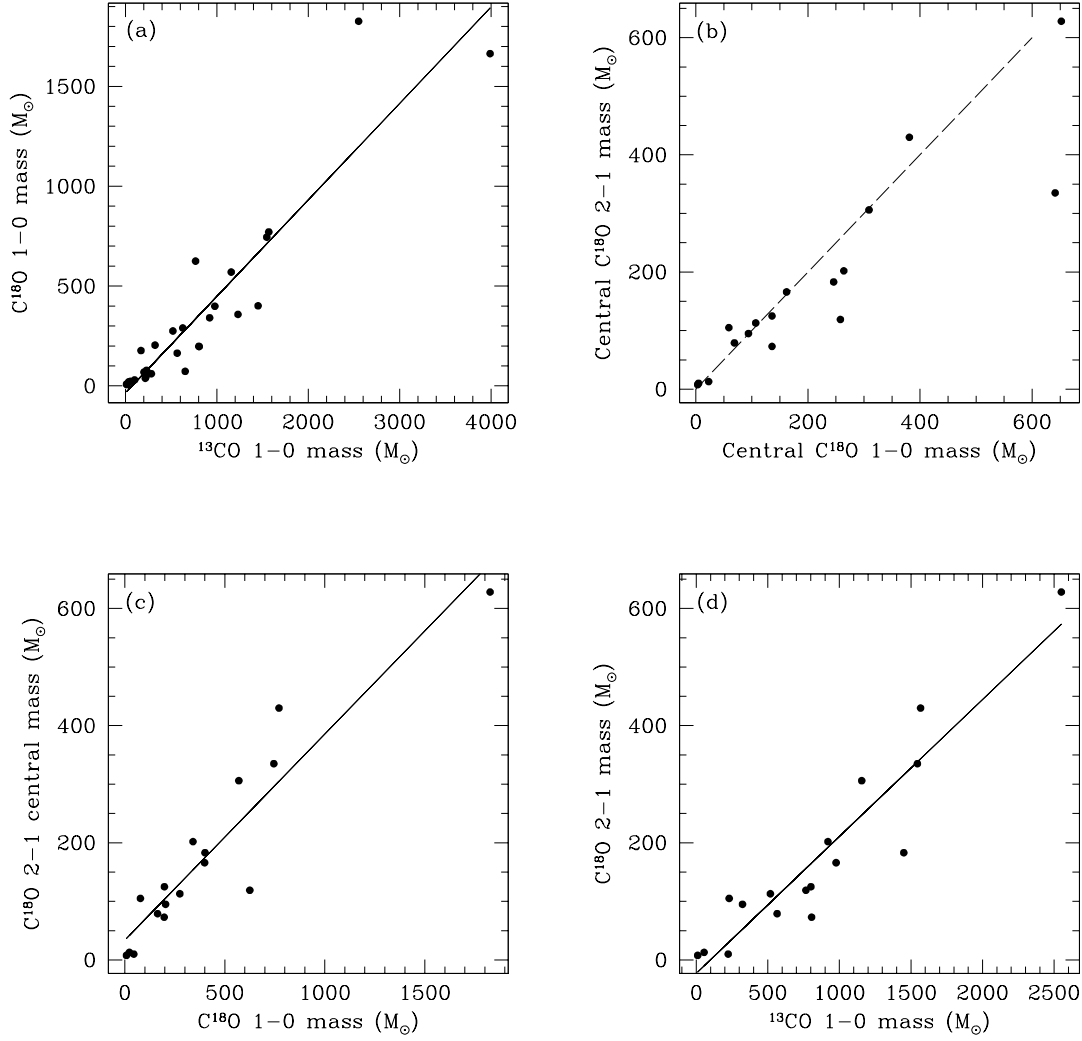}
\caption{(a) \xxco\ \hbox{1--0} mass vs. \xco\ \hbox{1--0} mass. Solid line 
indicates the best-fit line with a slope of 0.48$\pm$0.04, indicating
that the masses determined from the \xco\ observations are
approximately a factor of two larger.  
(b) Central \xxco\ \hbox{2--1} mass vs. Central \xxco\ \hbox{1--0}
mass. The dashed line indicates a 1:1 relationship. The two most
discrepant sources, GGD\,12-15 and CepC are marked.
(c) Central \xxco\ \hbox{2--1} mass vs. \xxco\ \hbox{1--0} mass. The
solid line indicates a best-fit line, with a slope of 0.35$\pm$0.03.
(d) Central mass from \xxco\ \hbox{2--1} emission vs. 'total' mass 
from the extended \xco\ \hbox{1--0} emission. The solid line 
indicates the best-fit line with a slope of 0.23$\pm$0.02.
\label{mass1}}
\end{figure}

%\onecolumn
\begin{figure}
\epsscale{}
\plottwo{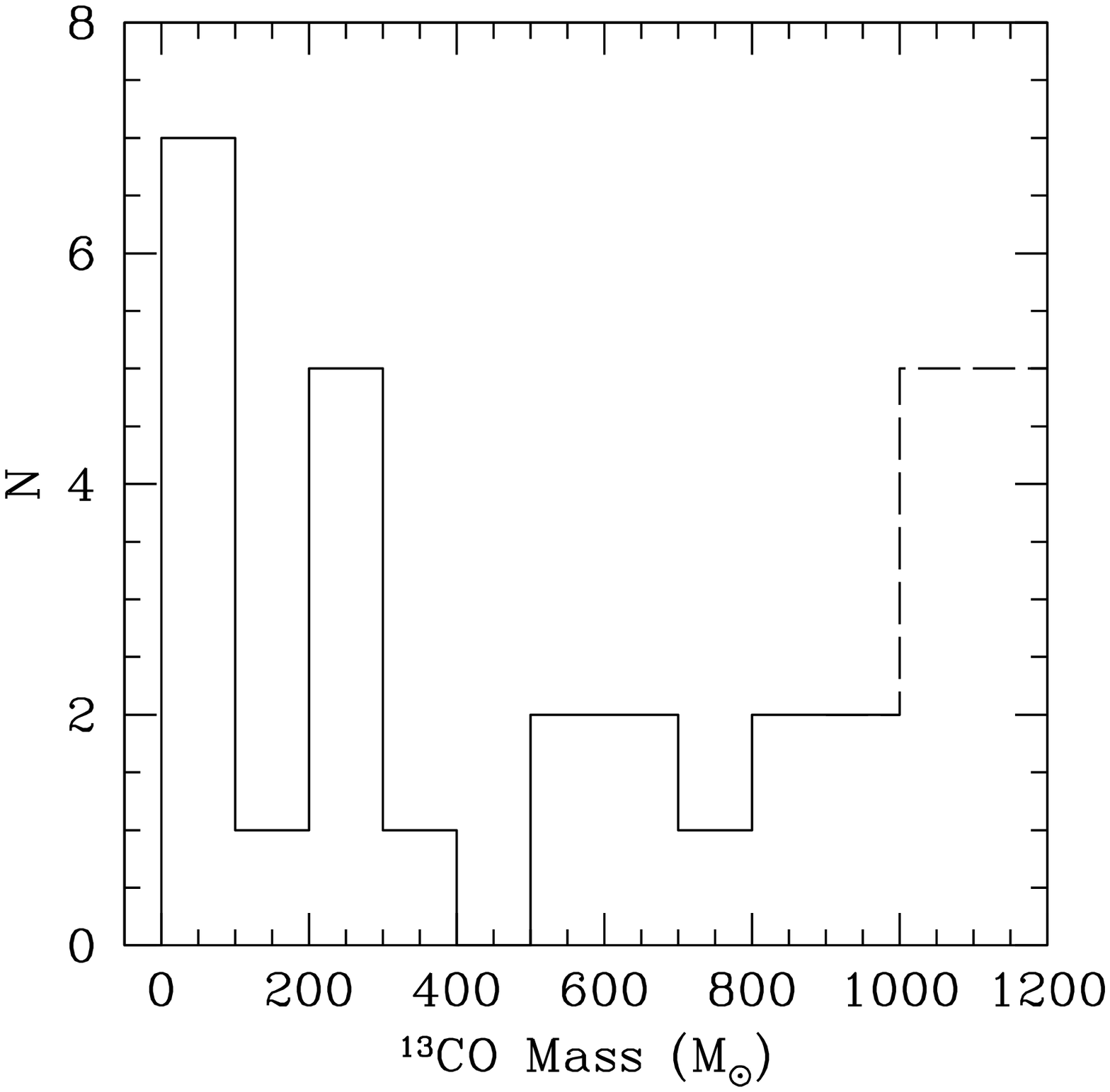}{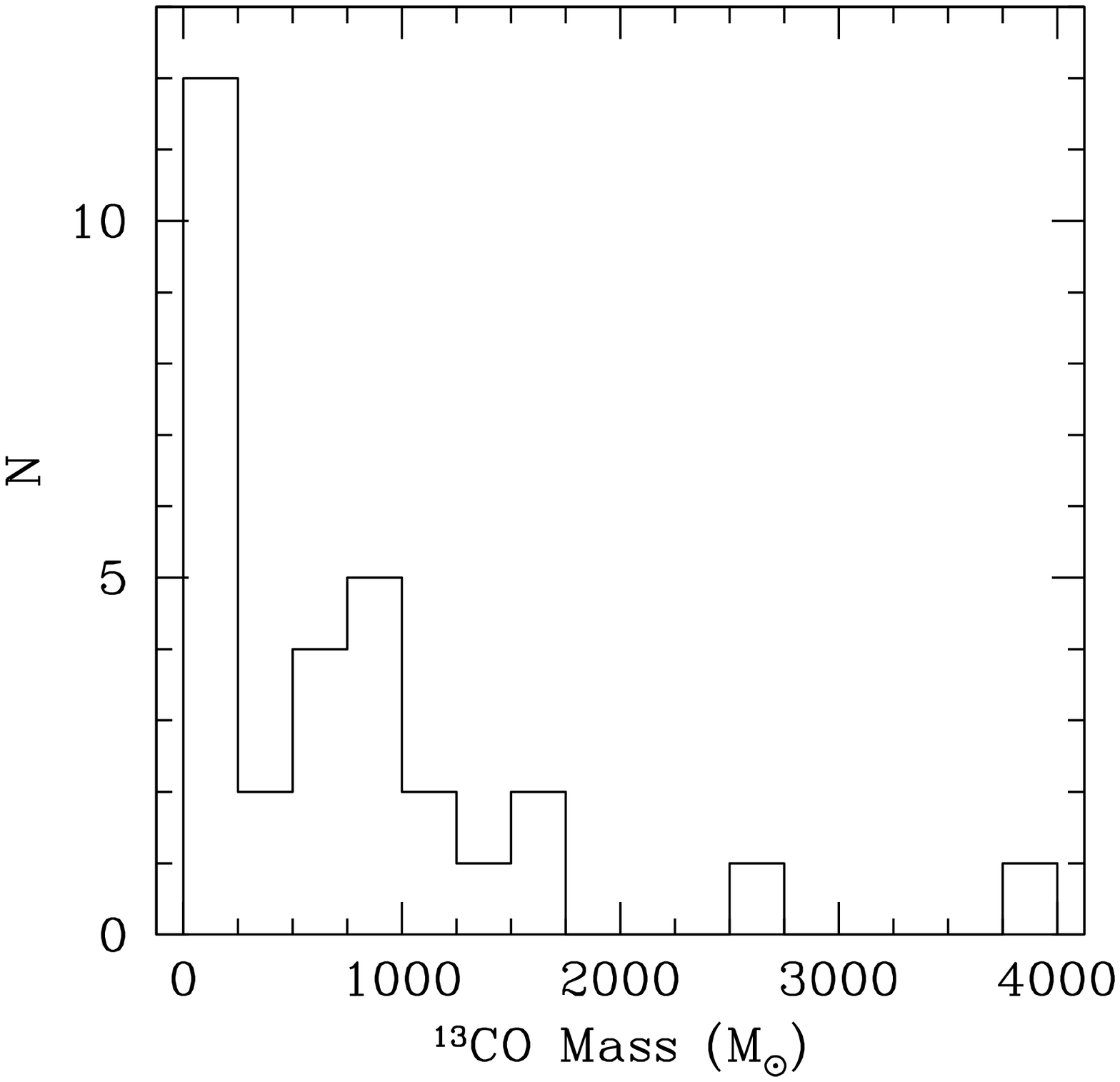}
\caption{(a) Histograms of \xco\ \hbox{1--0} masses. Left: Expanded scale
between 0 and 1000\Msolar\ with bin size of 100\Msolar. Right: All
sources with bin size of 250\Msolar.
\label{mass_hist}}
\end{figure}

%\onecolumn
\begin{figure}
\plottwo{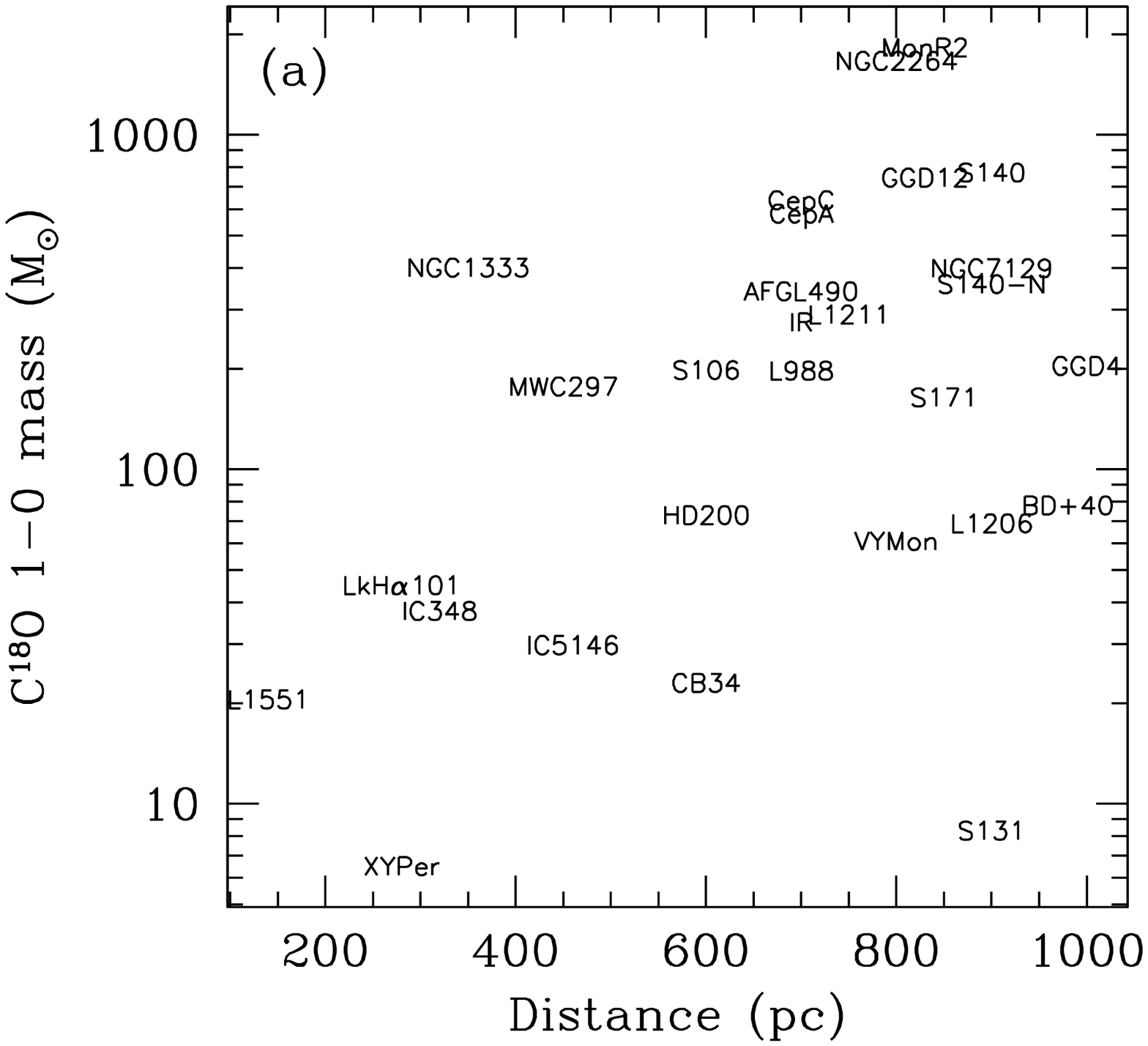}{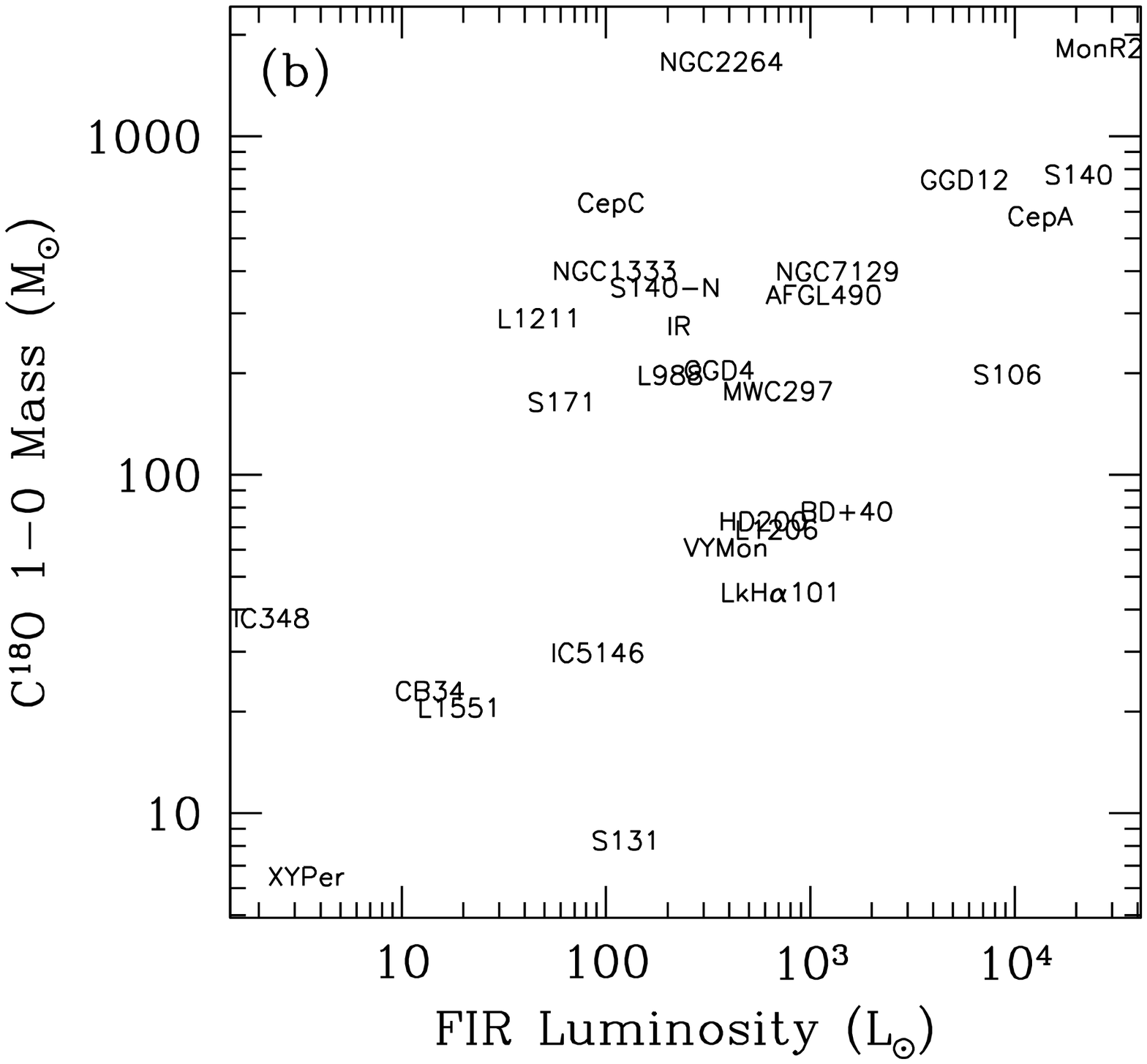}
\caption{(a) \xxco\ 1--0 mass vs. distance. Some source names have been shortened for clarity:
HD200 = HD\,200775; IR = IRAS\,20050; BD+40 = BD+40\degrees4124; L988 = L\,988-e; 
GGD12 = GGD\,12-15.
(b) \xxco\ 1--0 mass vs. FIR luminosity. Source abbreviations as for
(a).
\label{mass_plots}}
\end{figure}

\onecolumn
\begin{figure}
\epsscale{}
\plottwo{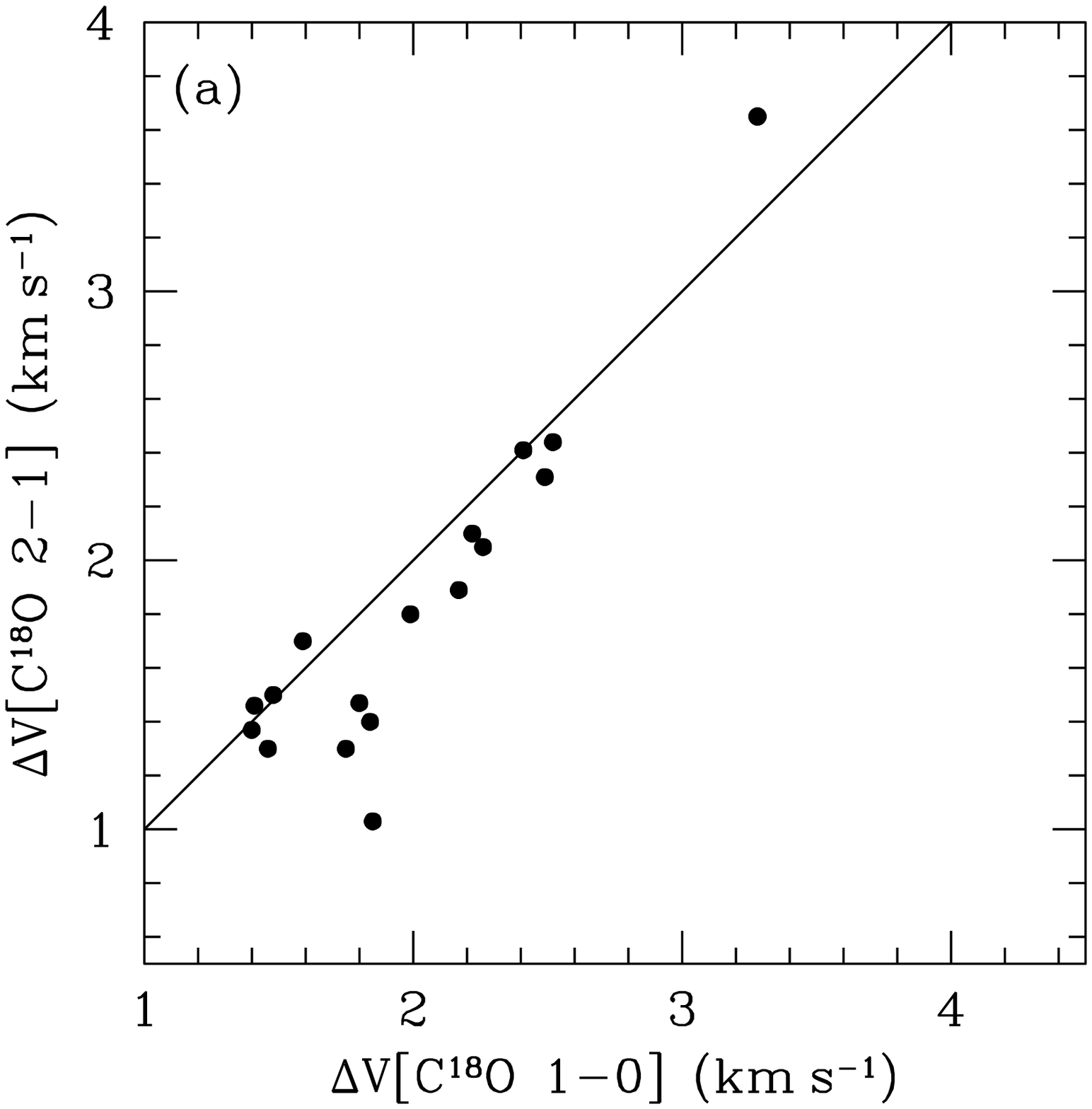}{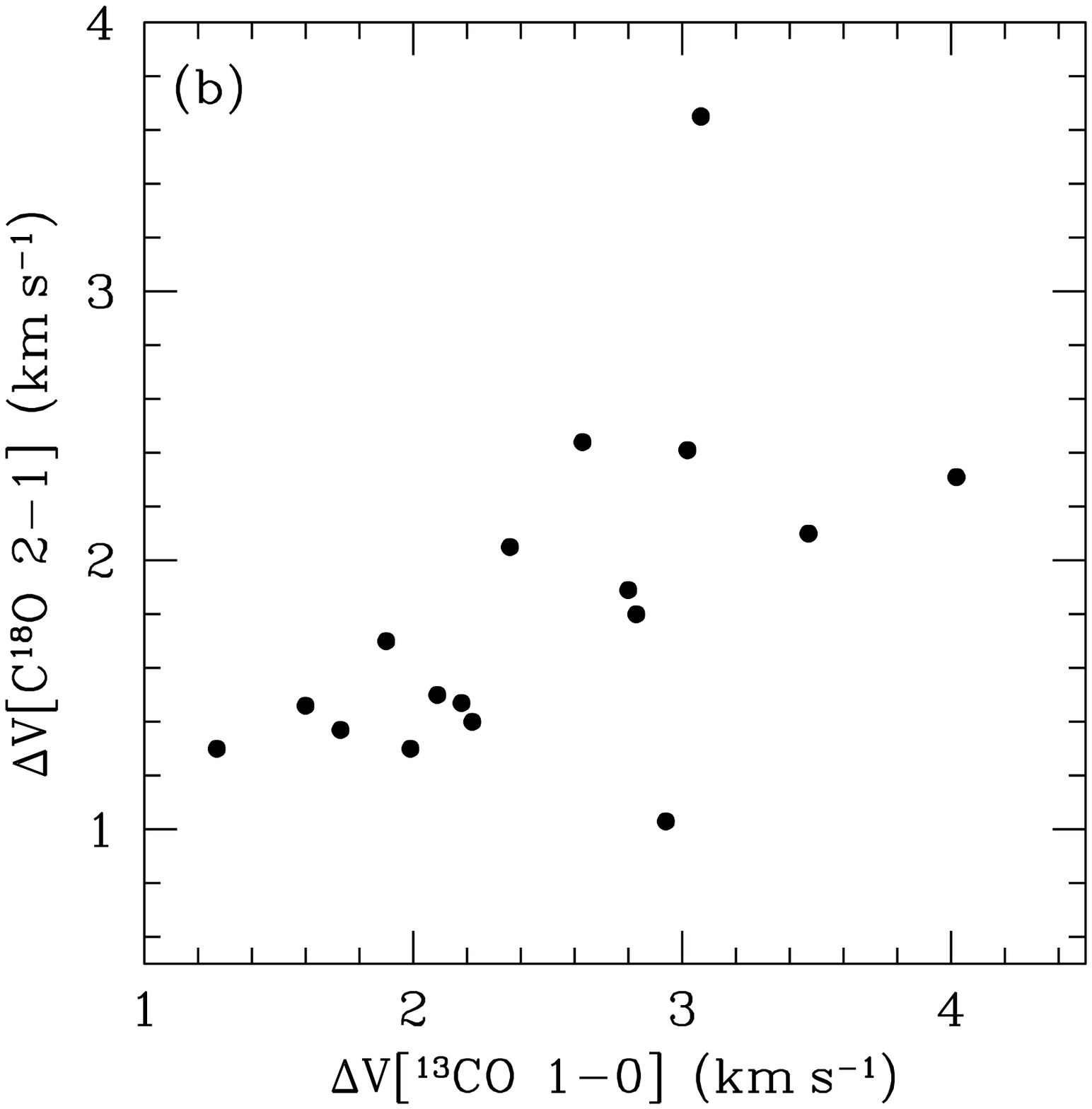}
\caption{(a) Average \xxco\ \hbox{2--1} linewidth vs.  average \xxco\ \hbox{1--0} linewidth. 
Solid line indicates a 1:1  relation. (b) Average \xxco\ \hbox{2--1} linewidth vs. average \xco\
\hbox{1--0} linewidth. The larger scatter is most likely due to
optical depth effects in the \xco\ spectra.
\label{lw}}
\end{figure}

\twocolumn
\begin{figure}
\plotone{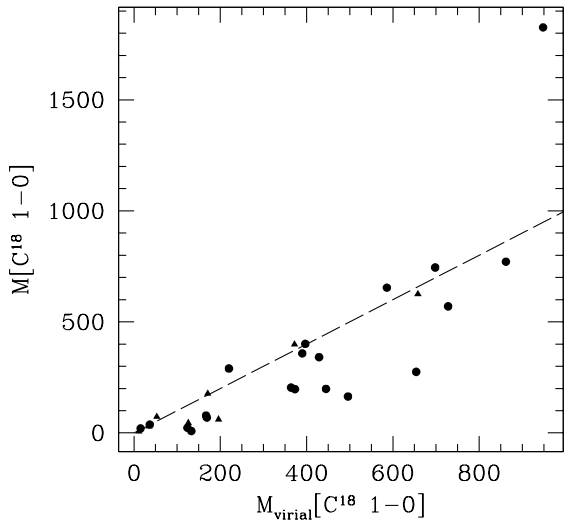}
\caption{\xxco~1--0 mass vs. \xxco~1--0 virial mass. 
Circles represent sources with centrally condensed morphologies, while
triangles are sources which are extended or diffuse (for which the
virial mass is not really applicable). The dashed line indicates the
1:1 relation.
\label{virial}}
\end{figure}

\onecolumn
\begin{figure}
\plottwo{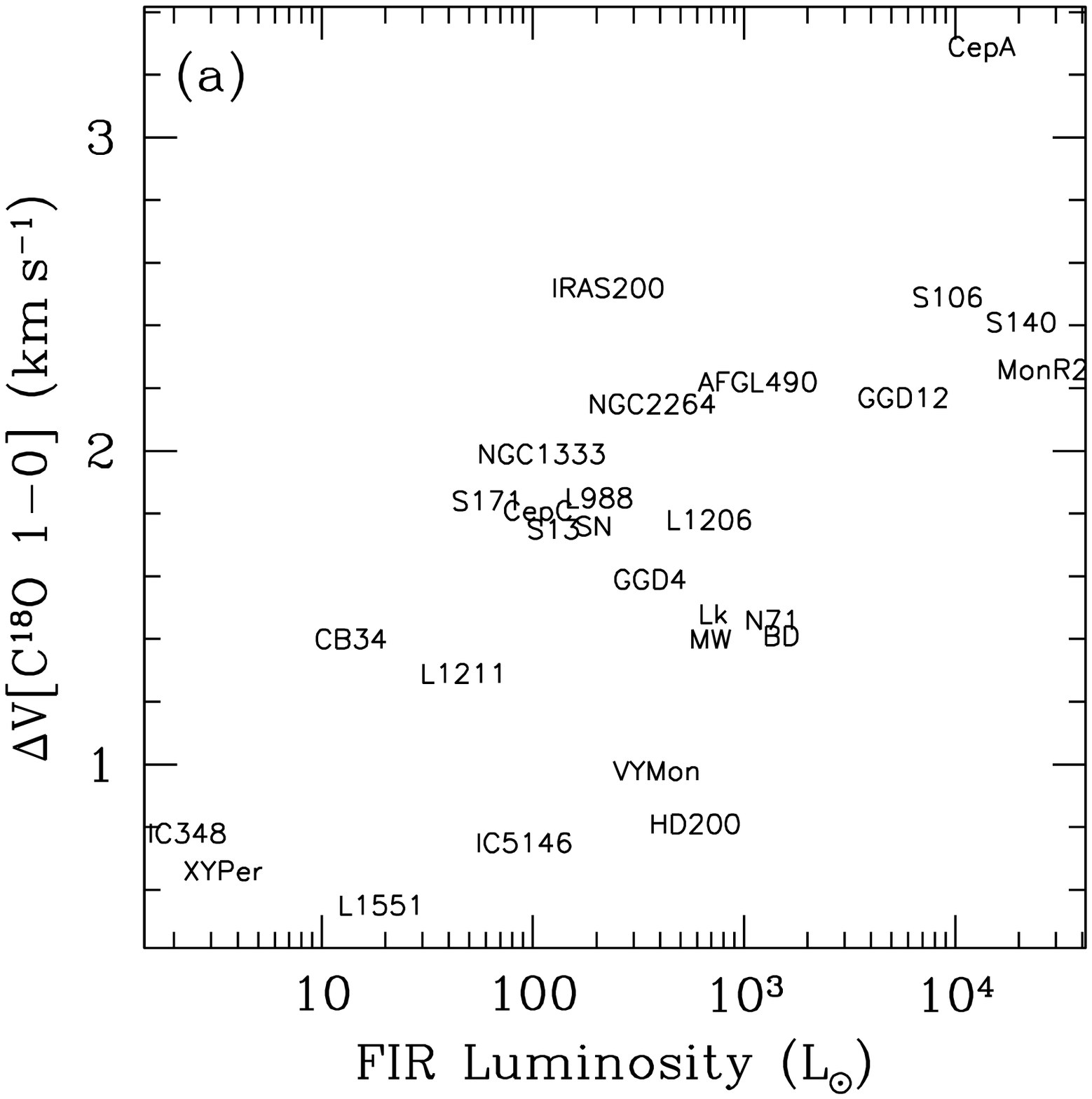}{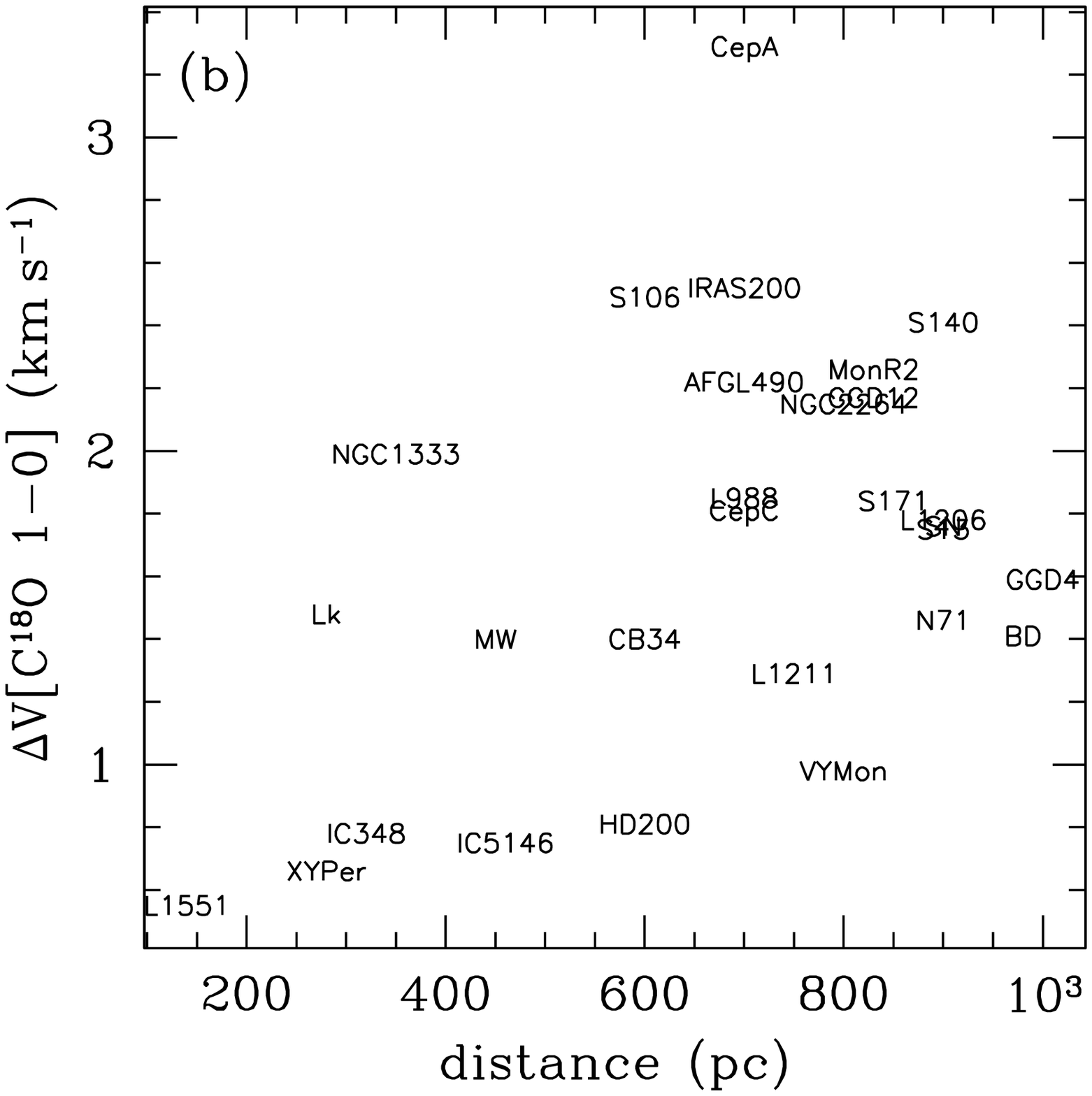}
\caption{(a) Average \xxco~1--0 linewidth vs. FIR luminosity.
Some source names have been abbreviated for clarity: HD200 = HD\,200775;
IRAS200 = IRAS\,20050; BD = BD+40\degrees4124;
N71 = NGC7129; Lk = LkH$\alpha$101; MW = MWC\,297; SN = S\,140-N;
S13 = S\,131.
(b)~Average \xxco~1--0 linewidth vs. distance.
\label{fir_lw}}
\end{figure}

\twocolumn
\begin{figure}
\plotone{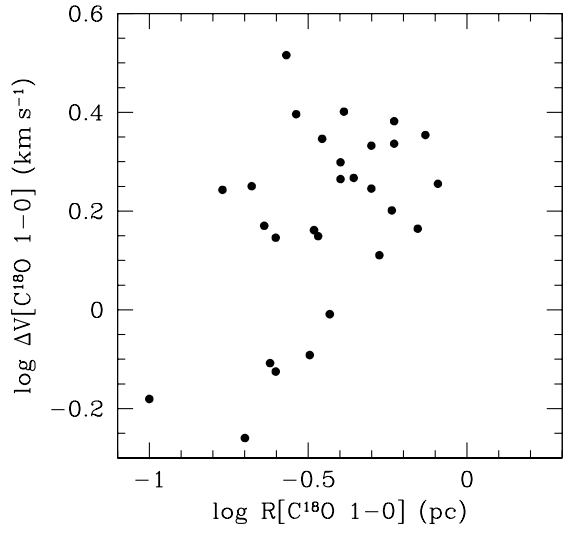}
\caption{Size--linewidth relation for the \xxco\ 1--0 line.
\label{size_lw}}
\end{figure}

\onecolumn
%%%%%%%%%%%%%%%%%%%%%%%%%%%%%%%%%%%%%%%%%%%%%%%%%%%%%%%%%%%%%%%%%%%%%%%%%%%%%%%
%% TABLES

\newpage
\begin{deluxetable}{lcccccc}
\tabletypesize{\scriptsize}
\tablecaption{The Sample\label{sourcetab}}
\tablewidth{0pt}
\tablehead{
\colhead{Source} & \colhead{RA(1950)}& \colhead{Dec(1950)}& \colhead{Adopted Distance}&\colhead{FIR Luminosity\tablenotemark{a}} &\colhead{N$_{stars}$}& \colhead{Refs.}\\
& \colhead{hh:mm:ss}& \colhead{dd:mm:ss}& \colhead{pc}&
\colhead{\Lsolar}&&\\ }
\startdata
\objectname{MWC297}       &  18:25:01.40 & $-$03:51:47.0  & 450 &694.2&37&1,2\\%%
\objectname{VVSer}        &  18:26:14.30 & +00:06:40.0  & 440 &10.7&24&1,2\\	%%
\objectname{IRAS20050}    &  20:05:02.00 & +27:20:30.0  & 700 &227.1&$\sim$100&3\\%%
\objectname{BD+40\degrees4124}&20:18:43.32 & +41:12:11.6 & 980 &1503&$\ga$19&4\\%%
\objectname{S106}         &  20:25:34.00 & +37:12:50.0  & 600 &9205&160&5\\	%%
\objectname{HD200775}     &  21:00:59.70 & +67:57:56.0  & 600 &585.9&9&1,2\\	%%
\objectname{L988-e}       &  21:02:04.60 & +50:03:20.0  & 700 &206.1&45&6\\	%%
\objectname{S131}         &  21:38:53.20 & +56:22:18.0  & 900 &125.1&$\la$10&7,8\\%%
\objectname{NGC\,7129}    &  21:41:47.00 & +65:52:45.5  & 900 &1360&29&1,2\\%%
\objectname{IC5146}	  &  21:51:36.23 & +47:01:54.9  & 460 &90.8&$>$100&17,18\\ 
\objectname{S140}         &  22:17:41.10 & +63:03:41.6  & 900 &20560&$\la$16&9\\%%	
\objectname{S140-N}       &  22:17:51.10 & +63:17:50.0  & 900 &194.1&$>$10&9\\
\objectname{L1206}        &  22:27:12.20 & +63:58:21.0  & 900 &684.4&27&6\\%%	
\objectname{L1211}        &  22:45:23.30 & +61:46:07.0  & 750 &47.0&$<$245\tablenotemark{b}&6\\%%	
\objectname{HD216629}     &  22:51:18.40 & +61:52:46.0  & 700 &25.0&29&1,2\\%%	
\objectname{CepA}         &  22:54:20.20 & +61:45:55.0  & 700 &13320&$<$580\tablenotemark{b}&6\\%%	
\objectname{CepC}         &  23:03:45.60 & +62:13:49.0  & 700 &105.9&$<$110\tablenotemark{b}&6\\%%
\objectname{S171}         &  00:01:23.00 & +68:17:59.0  & 850 &60.8&$>$10&10\\%%	
\objectname{AFGL490}      &  03:23:39.00 & +58:36:33.0  & 700 &1167&45&6\\%%
\objectname{NGC1333}      &  03:25:54.98 & +31:08:24.1  & 350 &110.0&143&11\\%%
\objectname{IC348}	  &  03:41:13.14 & +32:00:52.1  & 320 &2.3  &$\sim$400&15,19,24\\
\objectname{XYPer}        &  03:46:17.40 & +38:49:50.0 & 280 &3.4&11&1,2,4,12\\%%
\objectname{LkH$\alpha$101}& 04:26:57.30 & +35:09:56.0  & 280 &713.5&$>$100&12,13,16\\%%
\objectname{L1551}        &  04:28:45.14 & +18:04:44.3  & 140 &19.1 &$\ga$15 & 20,21\\
\objectname{GGD4}         &  05:37:21.30 & +23:49:22.0  & 1000 &359.0&10&6\\%%	
\objectname{CB34}         &  05:44:02.40 & +20:59:22.0  & 600 &13.8&12&14\\%%
\objectname{MonR2}        &  06:05:20.00 & $-$06:22:30.0  & 830 &26010&371&15\\%%	
\objectname{GGD12-15}     &  06:08:24.50 & $-$06:11:12.0  & 830 &5682&134&15\\	
\objectname{VYMon}        &  06:28:21.00 & +10:28:15.0  & 800 &387.4& 25&1,2 \\%%
\objectname{NGC2264}      &  06:38:10.00 & +09:40:00.0  & 800 &367.2& 360 &22,23\\
\enddata
\tablecomments{Source details compiled from the literature}
\tablenotetext{a}{Determined from the 12--100\micron\ fluxes of the nearest IRAS point source, as described in section \ref{sample}.}
\tablenotetext{b}{Limits indicate the estimate was not corrected for foreground or background
contamination, hence the true number of stars in the cluster will be
much smaller than the number quoted here.}
\tablerefs{
1.~\citealt{testi98};
2.~\citealt*{testi99};
3.~\citealt{chen97};
4.~\citealt{testi97};
5.~\citealt{hr91};
6.~\citealt{hodapp94};
7.~\citealt{sugi91}; 
8.~\citealt*{sugi95};
9.~\citealt{evans89};
10.~\citealt{yf92};
11.~\citealt*{lal96};
12.~A.\ Wilson 2002, personal communication;
13.~\citealt*{bsk91};
14.~\citealt{ay95}; 
15.~\citealt{carp2000};
16.~\citealt{ab94};
17.~\citealt*{lal99};
18.~\citealt{herbig02};
19.~\citealt{carp02};
20.~\citealt*{frid97};
21.~\citealt{cark96};
22.~\citealt*{lyg93};
23.~\citealt{shvh97};
24.~\citealt{ll95}.
}
\end{deluxetable}
%%%%%%%%%%%%%%%%%%%%%%%%%%%%%%%%%%%%%%%%%%%%%%%%%%%%%%%%%%%%%%%%%%%%%%%%%%%
\newpage
\begin{deluxetable}{cccccccccc}
\tabletypesize{\scriptsize}
\rotate
\tablecaption{Summary of Observations\label{obstab}}
\tablewidth{0pt}
\tablehead{
\colhead{Source} & \multicolumn{4}{c}{FCRAO Observations}&&\multicolumn{4}{c}{SMTO Observations}\\ 
\cline{2-5} \cline{7-10}
\colhead{} &\colhead{Map Size}&\colhead{Spectral Resolution}&\colhead{RMS\tablenotemark{a,b}} &\colhead{Lines Observed}
&&\colhead{Map Size}&\colhead{Spectral Resolution} &\colhead{RMS\tablenotemark{a}}&\colhead{Lines Observed}\\
\colhead{} &\colhead{(arcmin)}&\colhead{(kHz)}&\colhead{(K)}&\colhead{}&&\colhead{(arcmin)}&\colhead{(kHz)}&\colhead{(K)}}
\startdata
{MWC297}       &15$\x$15&25 &0.22&\xco\ 1--0, \xxco\ 1--0& &\nodata&\nodata&\nodata&\nodata\\
{VVSer}        &15$\x$15&25 &0.25&\xco\ 1--0, \xxco\ 1--0& &\nodata&\nodata&\nodata&\nodata\\
{IRAS20050}    &15$\x$15&25&0.21&\xco\ 1--0, \xxco\ 1--0& &5$\x$5&250&0.20&\xxco\ 2--1\\
{BD+40\degrees4124}&15$\x$15&50&0.12&\xco\ 1--0, \xxco\ 1--0& &5$\x$5&250&0.14&\xxco\ 2--1\\
{S106}         &15$\x$15&25 &0.16&\xco\ 1--0, \xxco\ 1--0& &5$\x$5&250&0.13&\xxco\ 2--1\\
{HD200775}     &15$\x$15&25 &0.22&\xco\ 1--0, \xxco\ 1--0& &\nodata&\nodata&\nodata&\nodata\\
{L988-e}       &16$\x$16&25 &0.34&\xco\ 1--0, \xxco\ 1--0& &5$\x$5&250&0.16&\xxco\ 2--1\\
{S131}         & 8$\x$8&25 &0.24&\xco\ 1--0, \xxco\ 1--0& &3$\x$3&250&0.11&\xxco\ 2--1\\
{NGC7129}   &15$\x$15&25&0.17&\xco\ 1--0, \xxco\ 1--0& &5$\x$5&250&0.17&\xxco\ 2--1\\
{IC5146}    &30$\x$30&25 && \xco\ 1--0, \xxco\ 1--0& &\nodata&\nodata&\nodata&\nodata\\
{S140}         &19$\x$18&25 &0.20&\xco\ 1--0, \xxco\ 1--0& &7.5$\x$7.5&250&0.19&\xxco\ 2--1\\
{S140-N}       &15$\x$15&25 & &\xco\ 1--0, \xxco\ 1--0& &\nodata&\nodata&\nodata&\nodata\\
{L1206}        &15$\x$15&50 &0.22&\xco\ 1--0, \xxco\ 1--0& &\nodata&\nodata&\nodata&\nodata\\
{L1211}        &19$\x$18&25 &0.22&\xco\ 1--0, \xxco\ 1--0& &\nodata&\nodata&\nodata&\nodata\\
{HD216629}     &16$\x$16&25  &0.35&\xco\ 1--0, \xxco\ 1--0& &\nodata&\nodata&\nodata&\nodata\\
{CepA}         &15$\x$15&25 &0.22&\xco\ 1--0, \xxco\ 1--0& &5$\x$5&250&0.17&\xxco\ 2--1\\
{CepC}         &16$\x$16&25 &0.17&\xco\ 1--0, \xxco\ 1--0& &5$\x$5&250&0.17&\xxco\ 2--1\\
{S171}         &18$\x$28&25 &0.19&\xco\ 1--0, \xxco\ 1--0& &5$\x$5&250&0.11&\xxco\ 2--1\\
{AFGL490}      &15$\x$15&50 &0.18&\xco\ 1--0, \xxco\ 1--0&&7.5$\x$7.5&250&0.09&\xxco\ 2--1\\
{NGC1333}      & 33$\x$39&50 &0.14 &\xco\ 1--0, \xxco\ 1--0& &12.5$\x$12.5&250&0.17&\xxco\ 2--1\\
{IC348}        &30$\x$30&25 && \xco\ 1--0, \xxco\ 1--0& &\nodata&\nodata&\nodata&\nodata \\
{XYPer}        &15$\x$15&50 &0.28&\xco\ 1--0, \xxco\ 1--0&& \nodata&\nodata&\nodata&\nodata\\
{LkH$\alpha$101}& 15$\x$19&25 &0.17&\xco\ 1--0, \xxco\ 1--0& & 5$\x$5&250&0.13&\xxco\ 2--1\\
{L1551}        &30$\x$30&25 && \xco\ 1--0, \xxco\ 1--0& &\nodata&\nodata&\nodata&\nodata\\
{GGD4}         &15$\x$15&25 &0.20&\xco\ 1--0, \xxco\ 1--0& &5$\x$5&250&0.16&\xxco\ 2--1\\
{CB34}         &16$\x$16&25 &0.18&\xco\ 1--0, \xxco\ 1--0& &5$\x$5&250&0.17&\xxco\ 2--1\\
{MonR2}        &15$\x$15&25 &0.25&\xco\ 1--0, \xxco\ 1--0&&7.5$\x$7.5&250&0.11&\xxco\ 2--1\\
{GGD12-15}     &15$\x$15&25 &0.19&\xco\ 1--0, \xxco\ 1--0&&7.5$\x$7.5&250&0.14&\xxco\ 2--1\\
{VYMon}        &15$\x$15 & 50&0.22&\xco\ 1--0, \xxco\ 1--0& &\nodata&\nodata&\nodata&\nodata\\
{NGC2264}      &60$\x$60&25 && \xco\ 1--0, \xxco\ 1--0& &\nodata&\nodata&\nodata&\nodata\\\\
\enddata
%\tablecomments{Put caption here}
\tablenotetext{a}{Average rms per channel.}
\tablenotetext{b}{The noise in the \xco\ and \xxco\ \hbox{1--0} 
data is the same, as they were taken simultaneously through the same
receiver. The slight difference in system temperature between the two
frequencies is negligible, as is any difference in spectrometer
noise.}
\end{deluxetable}

%%%%%%%%%%%%%%%%%%%%%%%%%%%%%%%%%%%%%%%%%%%%%%%%%%%%%%%%%%%%%%%%%%%%%%%%%%%%%%%%%%%%%%%%%%%%%
%%MORPHOLOGY ETC TABLE
\newpage
\begin{deluxetable}{ccccc}
\tabletypesize{\scriptsize}
%\rotate
\tablecaption{Morphological Properties\label{classtab}}
\tablewidth{0pt}
\tablehead{
\colhead{Source} & \colhead{Morphology}& \colhead{N$_{peaks}$}&\colhead{d$_*$\tablenotemark{a}}& \colhead{Developmental}\\
&&&\colhead{(pc)}&\colhead{Type}
}
\startdata
MWC297    &diffuse, extended&	3&0.25\tablenotemark{b}&III\\
VVSer     &diffuse, extended&	5&0.25\tablenotemark{b}&III\\
IRAS20050 &core + envelope&	2&0&II\\
BD+40\degrees4124 &compact core&1&0&I$_{\rm BRG}$	\\
S106      &compact core&	2&0&II\\
HD200775  &cavity&	5&0.45\tablenotemark{b}&III\\
L988-e     &core + envelope&	1&0.50&III\\
S131       &compact core&	1&0&I$_{\rm BRG}$\\
NGC\,7129&core + cavity&1&0&II\\
IC5146 & extended & 7 &0.12 & III\\
S140     &compact core, cometary&	1&0&I$_{\rm BRG}$\\
S140-N	 &core + envelope       &2  &0.2 &II\\
L1206    &compact core&	1&0&I$_{\rm BRG}$\\
L1211    &core	+ envelope&3&0&II\\
HD216629 &compact core&	3&0.60&III\\
CepA     &compact core&	1&0&I\\
CepC     &extended ridge&2&0.30&	II\\
S171      &compact core&	2&0.60\tablenotemark{c}&II\\
AFGL490  &core + envelope&1&0&	I\\
NGC1333	  &core + envelope&	1&0.6&II\\
IC348    & core + ridge & 1 &0.75 & III\\
XYPer    &extended ridge&2&0.25	&III\\
LkH$\alpha$101&cavity&2&0.25\tablenotemark{b}&III\\
L1551     & extended & 2 &0.06 & II\\
GGD4      &compact core&	1&0&I\\
CB34      &compact core&	1&0&I$_{\rm BRG}$\\
MonR2    &core + envelope&1&0&I	\\
GGD12-15 &core + envelope&1&0&	I\\
VYMon     &extended ridge&	3&0&II\\
NGC2264 &core + envelope& 1 &0 &I
\enddata
\tablenotetext{a}{A zero in this column indicates the position of the IRAS source or star is
within 2 map pixels (50$''$ $\simeq$ 1 beamwidth) of the position of peak \xco\ emission.}
\tablenotetext{b}{IRAS source or star is located in or close to a cavity.}
\tablenotetext{c}{Although the IRAS source \emph{is} close to peak of \xxco\ emission.}
\end{deluxetable}

%%%%%%%%%%%%%%%%%%%%%%%%%%%%%%%%%%%%%%%%%%%%%%%%%%%%%%%%%%%%%%%%%%%%%%%%%%%%%%%%%%%%%%%%%%%%
%% MASS TABLE

\newpage
\begin{deluxetable}{cccccccc}
\tabletypesize{\scriptsize}
\rotate
\tablecaption{Masses\label{masstab}}
\tablewidth{0pt}
\tablehead{
\colhead{(1)} & \colhead{(2)}& \colhead{(3)}& \colhead{(4)}& \colhead{(5)}&\colhead{(6)}&\colhead{(7)}&\colhead{(8)}\\
\colhead{Source} &\colhead{M[\xco\ 1--0]}&\colhead{M$_{virial}$[\xco\ 1--0]}&\colhead{M[\xxco\ 1--0]}&\colhead{M$_{virial}$[\xxco\ 1--0]}&\colhead{M$_{central}$[\xxco\ 1--0]}&\colhead{M[\xxco\ 2--1]}&\colhead{M$_{virial}$[\xxco\ 2--1]}\\
&\colhead{\Msolar}&\colhead{\Msolar}&\colhead{\Msolar}&\colhead{\Msolar}&\colhead{\Msolar}&\colhead{\Msolar}&\colhead{\Msolar}
}
\startdata
MWC297    &169&	 613	&177	&171	&\nodata&\nodata&\nodata\\
VVSer     &72&	 1374	&18	&\nodata&\nodata&\nodata&\nodata\\
IRAS20050 &518&	 1057	&275	&654	&107	&113	&238\\
BD+40\degrees4124&230&264&78	&167	&59	&105	&123\\
S106      &801&	 1919	&198	&445	&136	&125	&320\\
HD200775  &653&	 341	&73	&53	&\nodata&\nodata&\nodata\\
L988-e     &806&1224	&197	&373	&136	&73	&48\\
S131       &10&	 78	&8.3	&133	&3.4	&7.8	&67\\
NGC\,7129&977&979&399	&372	&162	&166	&152\\
IC5146   &99.3 &110 & 29.8 &34.5 &\nodata&\nodata&\nodata\\
S140     &  1567&1293	&771	&862	&381	&430	&363\\
S140-N   &1230  &721 & 358 &390 &\nodata&\nodata&\nodata\\
L1206    &202&	 298	&69	&169	&\nodata&\nodata&\nodata\\
L1211    &  626&514	&290	&220	&\nodata&\nodata&\nodata\\
HD216629 &  47&	 212	&\nodata&\nodata&\nodata&\nodata&\nodata\\
CepA     &  1156&844	&570	&728	&309	&306	&632\\
CepC       &766&1204	&625	&658	&258	&119	&189\\
S171      &565&	 874	&164	&496	&69	&79	&147\\
AFGL490   & 920&1655	&341	&429	&264	&202	&308\\
NGC1333   &1450 &1397	&401	&397	&246	&183	&162\\
IC348     & 216 &113 &37.7 &36.6&\nodata&\nodata&\nodata\\
pXYPer     & 17&	 335	&6.5	&11	&\nodata&\nodata&\nodata\\
LkH$\alpha$101&223&361	&45	&126	&5	&10	&50\\
L1551     &34.2 &47 & 20.5 &15.3 &\nodata&\nodata&\nodata\\
GGD4      &323&	 598	&204	&364	&94	&95	&311\\
CB34      &55&	 204	&23	&124	&23	&13	&85\\
MonR2     & 2550&911	&1826	&948	&652	&628	&462\\
GGD12-15  & 1545&1234	&745	&698	&641	&335	&241\\
VYMon     &282&	 432	& 61	&196	&\nodata&\nodata&\nodata\\
NGC2264   &3988 &1517 & 1664 &586 &\nodata&\nodata&\nodata
\enddata
%\tablecomments{Put caption here}
\end{deluxetable}

%%%%%%%%%%%%%%%%%%%%%%%%%%%%%%%%%%%%%%%%%%%%%%%%%%%%%%%%%%%%%%%%%%%%%%%%%%%%%%%%%%%%%%%%%%%%
%%SIZE LINEWIDTH TABLE

\newpage
\begin{deluxetable}{ccccccc}
\tabletypesize{\scriptsize}
%\rotate
\tablecaption{Linewidths and Sizes \label{lwtab}}
\tablewidth{0pt}
\tablehead{
\colhead{(1)} & \colhead{(2)}& \colhead{(3)}& \colhead{(4)}& \colhead{(5)}&\colhead{(6)}&\colhead{(7)}\\
\colhead{Source} &\colhead{$\Delta$V[\xco\ 1--0]}&\colhead{R[\xco\ 1--0]}&\colhead{$\Delta$V[\xxco\ 1--0]}&\colhead{R[\xxco\ 1--0]}&\colhead{$\Delta$V[\xxco\ 2--1]}&\colhead{R[\xxco\ 2--1]}\\
&\colhead{\kms}&\colhead{pc}&\colhead{\kms}&\colhead{pc}&\colhead{\kms}&\colhead{pc}
}
\startdata
MWC297    &  1.89	  &0.68	  &1.45	  &0.33	  &\nodata&\nodata\\
VVSer     &  2.60	  &0.81	  &1.19	  &\nodata&\nodata&\nodata\\
IRAS20050 &  2.63	  &0.61	  &2.52	  &0.41	  &2.44	  &0.16	  \\
BD+40\degrees4124 &1.60	  &0.41	  &1.41	  &0.34	  &1.46   &0.23	  \\
S106      &  4.02	  &0.47	  &2.49	  &0.29	  &2.31	  &0.24	  \\
HD200775  &  1.51	  &0.60	  &0.81	  &0.32	  &\nodata&\nodata\\
L988-e     &  2.94	  &0.57	  &1.85	  &0.44	  &1.03	  &0.18	  \\
S131       &  1.27	  &0.19	  &1.75	  &0.17	  &1.30	  &0.16	  \\
NGC\,7129& 1.99	          &0.99	  &1.46	  &0.70	  &1.30	  &0.36	  \\
IC5146	 & 1.1		  &0.36   &0.75   &0.25	  &\nodata&\nodata\\
S140     &  3.02	  &0.57	  &2.41	  &0.59	  &2.41	  &0.25	  \\
S140-N   & 1.87           &0.83   &1.76   &0.50   &\nodata&\nodata\\
L1206    &  1.70	  &0.41	  &1.78	  &0.21	  &\nodata&\nodata\\
L1211    &  1.61	  &0.79	  &1.29	  &0.53	  &\nodata&\nodata\\
HD216629 &  1.60	  &0.33	  &\nodata&\nodata&\nodata&\nodata\\
CepA     &  3.07	  &0.36	  &3.28	  &0.27	  &3.65	  &0.19	  \\
CepC      &  2.18	  &1.01	  &1.80	  &0.81	  &1.47	  &0.35	  \\
S171      &  2.22	  &0.71	  &1.84	  &0.40	  &1.40	  &0.30	  \\
AFGL490   &  3.47	  &0.55	  &2.22	  &0.35	  &2.10	  &0.28	  \\
NGC1333	  &  2.83	  &0.69	  &1.99	  &0.40	  &1.80	  &0.20	  \\
IC348	  &  1.3          &0.27   &0.78   &0.24   &\nodata&\nodata\\
XYPer     &  2.16	  &0.29	  &0.66	  &0.10	  &\nodata&\nodata\\
LkH$\alpha$101 &2.09	  &0.33	  &1.48	  &0.23	  &1.50	  &0.09	  \\
L1551     &  0.79         &0.30   &0.55   &0.20   &\nodata&\nodata\\
GGD4      &  1.90	  &0.66	  &1.59	  &0.58	  &1.70	  &0.43	  \\
CB34      &  1.73	  &0.27	  &1.40	  &0.25	  &1.37	  &0.18	  \\
MonR2     &  2.36	  &0.65	  &2.26	  &0.74	  &2.05	  &0.44	  \\
GGD12-15  &  2.80	  &0.63	  &2.17	  &0.59	  &1.89	  &0.27	  \\
VYMon     &  1.46	  &0.81	  &0.98	  &0.37	  &\nodata&\nodata\\
NGC2264   &  2.72         &0.82   &2.15   &0.50   &\nodata&\nodata
\enddata
%\tablecomments{Put caption here}
\end{deluxetable}

%%---------------------------------------------------------------------------------

\end{document}